\let\ifpdf\relax
\documentclass{JINST}
\let\ifpdf\relax
\usepackage{graphicx}
\usepackage{booktabs}
\usepackage{array}
\usepackage{subfig}
\usepackage{url}
\usepackage{float}

\title{ Production of  gamma rays by pulsed laser beam Compton scattering off GeV-electrons using a non-planar four-mirror optical cavity}
\author{T. Akagi$^{e}$, S. Araki$^d$, J. Bonis$^a$, I. Chaikovska$^a$\thanks{Corresponding author.}, R. Chiche$^a$, R. Cizeron$^a$, M. Cohen$^a$, E. Cormier$^b$, P. Cornebise$^a$, N. Delerue$^a$, R. Flaminio$^c$, S. Funahashi$^d$, D.~Jehanno$^a$, Y. Honda$^d$, F. Labaye$^{a}$, M. Lacroix$^a$, R. Marie$^a$, C. Michel$^c$, S. Miyoshi$^{e}$,  S.~Nagata$^e$, T.~Omori$^d$, Y.~Peinaud$^a$,  L. Pinard$^c$, H. Shimizu$^d$, V. Soskov$^a$, T.~Takahashi$^e$,  R.~Tanaka$^e$, T.~Terunuma$^d$, J.~Urakawa$^d$,  A. Variola$^a$, F. Zomer$^a$.\\
\llap{$^a$}LAL, CNRS-IN2P3, Universit\'e Paris-Sud XI\\
 Centre Scientifique d'Orsay, B\^atiment 200, BP 34,  91898 Orsay cedex , France\\
\llap{$^b$}CELIA, CNRS, Universit\'e de Bordeaux \\
Domaine du Haut Carr\'e,  43, Rue Pierre Noailles, 33405 Talence, France\\
\llap{$^c$}LMA, CNRS-IN2P3, Universit\'e Claude Bernard Lyon I \\
B\^atiment Virgo 7, Avenue Pierre de Coubertin,  69622 Villeurbanne  cedex, France\\
\llap{$^d$}ATF, High Energy Accelerator Research Organization (KEK) \\
1-1 Oho, Tsukuba, 305-0801 Ibaraki,  Japan\\
\llap{$^e$}Hiroshima University \\
1-3-2 Kagamiyama, Higashi-Hiroshima 739-8511, Japan \\
  E-mail: \email{chaikovs@lal.in2p3.fr}} 

\abstract{As part of the positron source R\&D for future $e^+-e^-$ colliders and Compton based compact light sources, a high finesse non-planar four-mirror Fabry-Perot  cavity has recently been installed at the ATF~(KEK, Tsukuba, Japan)~\cite{intrumentationPaper}. The first measurements of the gamma ray flux produced with a such cavity using a pulsed laser is presented here. We demonstrate the production of a flux of 2.7~$\pm$~0.2~gamma rays per bunch crossing ($\sim3\times10^6$~gammas per second) during the commissioning.}

\keywords{Positrons production; Compton scattering; Fabry-Perot cavity; Gamma rays; ATF}

\begin{document}

\section{Introduction}

High Energy physics~\cite{araki2005design,ILCpositrons,CLICpositrons} as well as applied physics~\cite{louvre,ThomX,ThomX2,quantumbeam} are showing a great interest  for intense flux of high energy X-rays and gamma rays.

These gamma or alternatively X-rays can be generated by Compton scattering~\cite{compactref}. To achieve  a high flux of high energy photons despite the low cross section of Compton scattering one requires a high average power laser system based on a Fabry-Perot cavity~\cite{kogel} together with a high current electron beam~\cite{ruth}.
In this context, a two-mirror  Fabry-Perot cavity has already been successfully operated at the Accelerator Test Facility~\cite{ATF} (ATF)  of KEK~\cite{shimizu2009photon,miyoshi2010photon}.

A solution to produce a high flux of circularly polarized gamma rays  is to use a four-mirror Fabry-Perot cavity (FPC) where laser pulses can be stacked to reach a high average power at the interaction point (IP). A prototype of non-planar high finesse four-mirror Fabry-Perot cavity has been installed at the KEK ATF and is described in details in~\cite{intrumentationPaper}. The optical system has been commissioned during summer 2010 and electron-photon Compton collisions were observed on the first attempt in October 2010. In this paper we present measurements of the gamma ray flux recorded during our commissioning. The data analyzed here were taken before the tragic earthquake which struck Japan in March 2011. Recovery and improvement work is being carried out since but is not covered by this paper. 

This paper is organised as follows. Section~\ref{expsetup} describes the experimental setup used to produce and measure the gamma rays. Section~\ref{sec:expectgflux} discusses the gamma ray flux expected with our current setup. Section~\ref{sec:measur} explains how the data were analyzed. Section~\ref{sec:results} presents the gamma ray spectra measured. Finally, the main results and further steps are discussed in section~\ref{sec:concl}.

\section{Experimental setup}
\label{expsetup}
\subsection{ Accelerator Test Facility}

\begin{figure}[htbp]
\begin{center}
\includegraphics[width=0.9\textwidth]{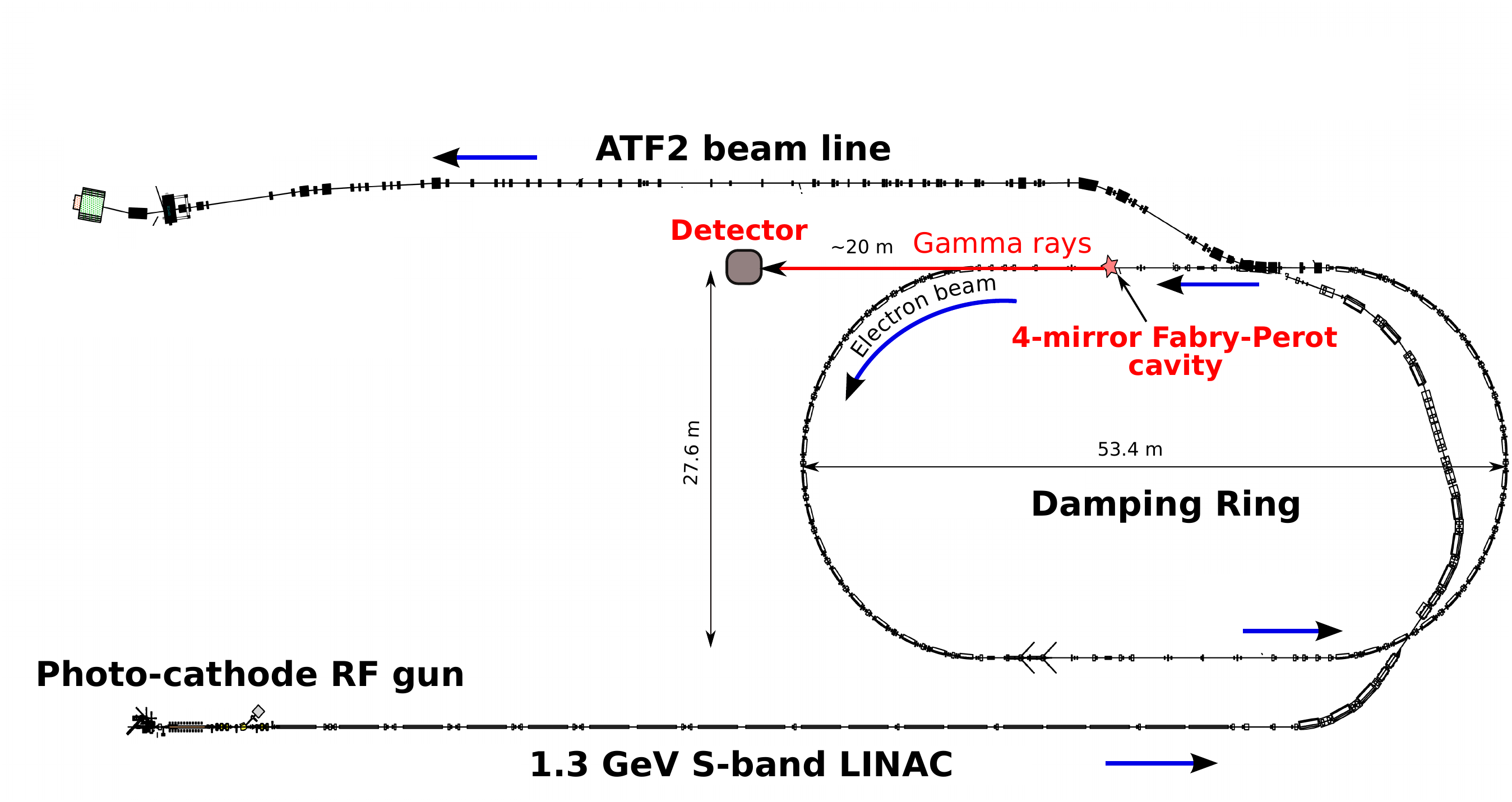}
\caption{Layout of the Accelerator Test Facility at KEK. The red star indicates the approximate location of  the 4-mirror Fabry-Perot cavity described in~\cite{intrumentationPaper} and used for the measurements presented in this paper. The red arrow indicates the direction of propagation of the  gamma rays produced. The grey box indicates the gamma ray detector (adapted from~\cite{miyoshi2010photon}).}
\label{fig:ATF_DR}
\end{center}
\end{figure}

A detailed description of the ATF at KEK can be found in~\cite{ATF,ATF-2000,ATF2}. The 1.28~GeV damping ring~(DR) has a revolution period of 462~ns and operates at the radio frequency (RF) of 357~MHz (165~RF~buckets spaced by 2.8ns). Although a total of up to 3 trains of 10 bunches separated by 5.6~ns can be injected in the ring, most operations run with a single bunch in the train or with a train of up to 10 bunches~\cite{ATF-Timing}.  Our FPC is installed in one of the straight sections of the damping ring as shown on figure~\ref{fig:ATF_DR}.  After the collisions the gamma rays, propagating along the electron beam, are extracted through a window before passing through several collimators. They are measured by a gamma detector about 20~m downstream the IP.

The main parameters of the electron beam at the ATF DR are given in the table~\ref{table:parameters}. 

\begin{table}[h]
\caption{ATF DR parameters.}
\begin{tabular}{b{0.7\textwidth}m{0.3\textwidth}}
\toprule
\textbf{Description} & \textbf{Value} \\ 
\toprule
Electron energy, $E_e$         & 1.28\,GeV \\
Electron bunch population, $N_e$      & $\sim$~0.5 $\times 10^{10}$ \\
Electron bunch length, $\sigma_z / c = \tau_{e}$  &$\sim$~20\, ps \\
Electron beam size, $\sigma_x / \sigma_y$  & $\sim$~110 / 10\, $\mu m$\\
Revolution period, $T_0$ & 462\, ns \\
Emittance, $\gamma \epsilon_{x/y}$  & 5$\times 10^{-6}$ / 3$\times 10^{-8}$ \, m$\cdot$rad \\
 \bottomrule
 \end{tabular}
\label{table:parameters}
\end{table}

\subsection{Fabry-Perot cavity}
The FPC is composed of 2 concave mirrors with a radius of curvature of 0.5~m and two flat mirrors with a non-planar tetrahedron geometry. The mirrors have a very high reflectivity (1 - 1060~ppm for one of them and 1 - 330~ppm for the others) leading to a cavity finesse of the order of 3000 (corresponding to a passive power enhancement of about 1000). A round trip time of 5.6~ns in the FPC is set to match half the RF frequency of the ATF DR at 178.5~MHz.

A passively mode-locked oscillator is amplified in a microstructured active fiber and further injected in the FPC. An all-digital double feedback system is used to lock the cavity on the ATF clock and to ensure that the laser pulses are properly stacked with interferometric accuracy.

The FPC is mounted on an optical table which is itself mounted on movers that allow to control precisely the vertical position of the FPC with respect to that of the laser beam.

Table~\ref{table:LaserParameters} summarises the main parameters of the system and more details on the experimental apparatus can be found in~\cite{intrumentationPaper}. 

\begin{table}[h]
\caption{Parameters of the laser system used during the data taking presented here.}
\begin{tabular}{b{0.7\textwidth}m{0.3\textwidth}}
\toprule
\textbf{Description} & \textbf{Value} \\ 
\toprule
Laser photon energy, $E_{ph}$  & 1.2\, eV  ($\lambda_{ph} = $1032 nm) \\
Laser spot size, $\sigma_x / \sigma_y$  & 26 / 38\, $\mu m$\\
Laser repetition rate, $f_L$  & 178.5\, MHz $\pm$ 4 KHz \\
Finesse, $F$  & $\sim$~3000  \\
Laser pulse length, $\tau_L$  & 68\, ps  \\
Average power stored in FPC, $P_L$  & $\sim$~160\, W  \\
Crossing angle, $\phi$  & 8\, deg.  \\
 \bottomrule
 \end{tabular}
\label{table:LaserParameters}
\end{table}

As there is an odd number of RF buckets in the DR collisions on a given electron bunch occur only every other turn, at a frequency of 1.08~MHz.

\paragraph{Data acquisition}


The data acquisition relies on several oscilloscopes to acquire information from the FPC and from the calorimeter. The waveforms acquired are timestamped in a database and saved on disk for later analysis. Data from the accelerator such as beam position and charge are also acquired and saved in the database.




\subsection{Calorimeter}

To detect the gamma rays produced by the collisions between the laser photons and  the electrons, we use a fast scintillation detector made of  barium fluoride ($BaF_2$) coupled with a Photomultiplier Tube~(PMT). Two polished rectangular crystals of $BaF_2$ with dimensions 100~mm~$\times$~70~mm~$\times$~70~mm are glued to each other and wrapped into an absorbing tape. 
The 200~mm depth of the $BaF_2$ calorimeter is nearly 10 radiation lengths, ensuring high detection efficiency~\cite{Miyoshithesis}.

Barium fluoride is commonly used as it has two emission spectra peaks at 220~nm and 310~nm with decay time constants of about 0.8~ns and 630~ns respectively. 
The decay time of the fast component of $BaF_2$ allows to resolve two successive pulses of gamma rays generated from the electron bunches spaced by 5.6~ns. 
To eliminate the slow component of the scintillation an optical filter has been installed in front of the PMT. A fast PMT (Hamamatsu Photonics R3377) with a rise time of 0.7 ns is used. Data acquisition is performed using a LeCroy WS454 oscilloscope (1GS/s, 500 MHz bandwidth). 

\paragraph{GEANT4 simulations}
Using Geant4~\cite{GEANT4} we simulated the detector to study the development of electromagnetic showers inside the $BaF_2$ calorimeter.

This allows to perform realistic simulations of high energy gammas passing through the calorimeter.  The initial distribution of the gamma rays is obtained by using the beam-beam interaction code CAIN2.40~\cite{cain}. Once a gamma ray hits the calorimeter, scintillation light is uniformly emitted along the path of the charged particles produced by the electromagnetic shower. The number of optical photons generated is proportional to the energy loss of the initial gamma rays. The scintillation light then propagates and finally reaches the PMT located at the end of the scintillator.

\section{Expected gamma ray flux}
\label{sec:expectgflux}
Using the Compton scattering physical properties it is possible to estimate the expected number of scattered gamma rays in the experiment. Neglecting the divergence of the electron and laser beam at the IP, the luminosity for the Compton collisions is given by equation~\ref{eq:1} in~\cite{suzukigeneral}:
\begin{equation}
\label{eq:1}
\mathcal{L} = N_eN_{ph}f \frac{\cos(\phi/2)}{2\pi}\frac{1}{\sqrt{\sigma_{ye}^2 + \sigma_{yph}^2 }\sqrt{(\sigma_{xph}^2 + \sigma_{xe}^2)\cos^2(\phi/2) + (\sigma_{ze}^2 + \sigma_{zph}^2)\sin^2(\phi/2) }},
\end{equation}
\begin{equation}
\label{eq:2}
\mathcal{F} =\frac{dN_{\gamma}}{dt}= \sigma_{Compton}\cdot \mathcal{L},
\end{equation}
where $N_e$, $N_{ph}$ are the number of the electrons in the bunch and number of the photons in the laser pulse respectively, $f$ is the repetition frequency, $\phi$ is the angle of the collisions, $\sigma_{\vec{r}e}$ and  $\sigma_{\vec{r}ph}$ indicate respectively the RMS sizes of the electron bunch and the laser pulse. Once, the luminosity of the process is defined, the flux of gamma rays is given by formula~\ref{eq:2}, where $ \sigma_{Compton}$ is the total Compton scattering cross section which is determined by the momenta of the incident electron and laser photon~\cite{landau}.

Formulas~\ref{eq:1} and~\ref{eq:2} show that the emitted rate is inversely proportional to the transverse electron and laser beam sizes. These formulas shows also that a collision angle reduces the gamma ray flux especially in the case of long electron bunches and laser pulses.

The simulation code CAIN has been used to simulate the Compton scattering process. This code simulates the interaction between one electron bunch and one laser pulse where the Compton cross section is calculated for all the possible polarization states~\cite{landau} and takes into account the beam geometry. 
Using the parameters listed in table~\ref{table:parameters} and~\ref{table:LaserParameters} the energy spectrum and expected number of the scattered gamma rays per bunch crossing have been obtained. 

\begin{figure}[htb]
   \centering
   \includegraphics[width=0.5\textwidth]{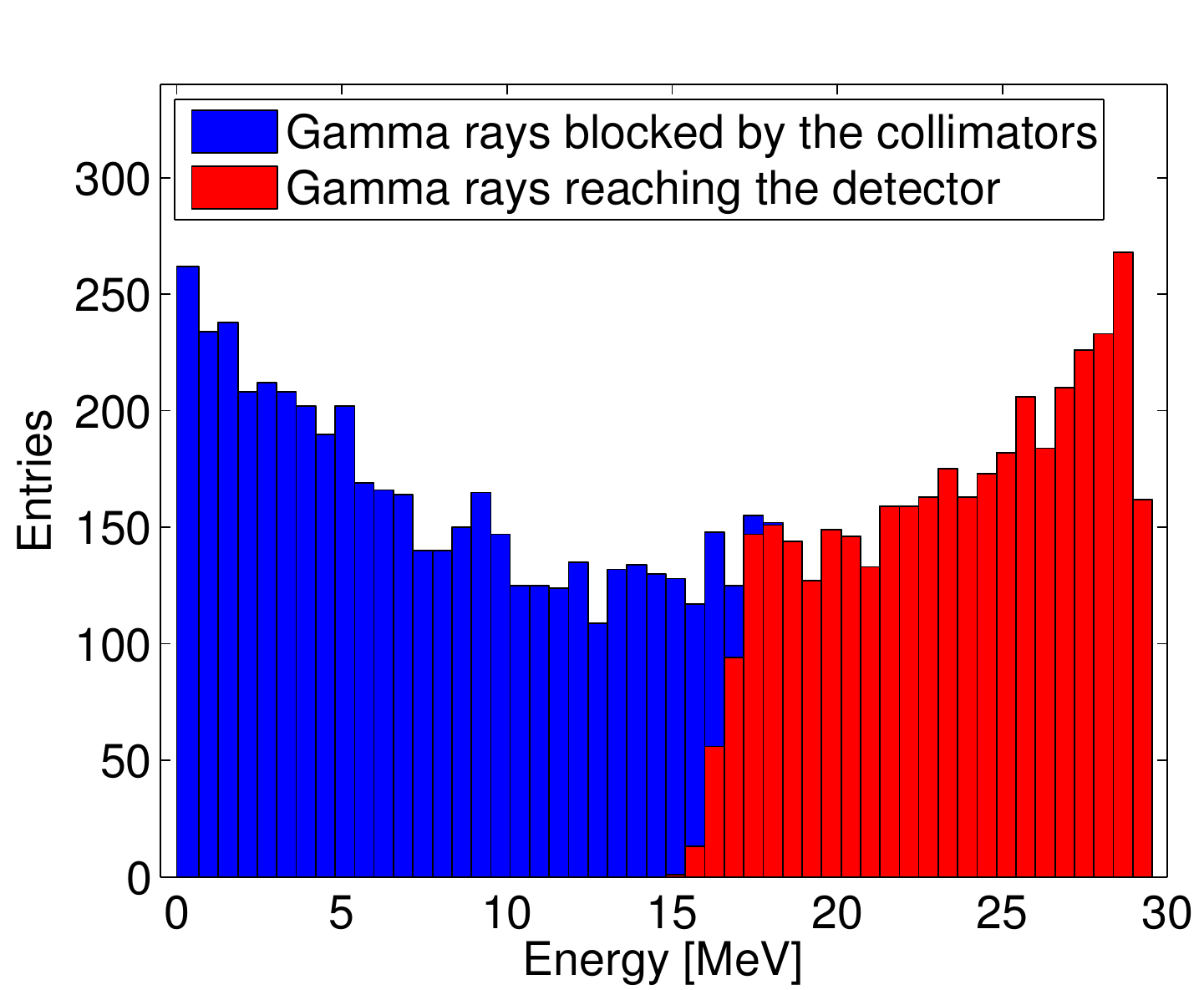}
   \caption{Energy spectrum of the gamma rays. The energies of the gamma rays blocked by the collimators are shown as well as the energies of the gamma rays entering the detector. Only the gamma rays with energies above 15~MeV are accepted by the collimators.}
   \label{fig:simu_spectrum}
\end{figure}

As it was mentioned before, in the experiment the scattered gamma rays pass through several collimators before entering the detector. This carries out a gamma energy spectrum selection owing to the energy-angle correlation in Compton scattering.
Figure~\ref{fig:simu_spectrum} shows the simulated energy spectrum of the gamma rays where the red color refers to the gamma rays transmitted by the collimators and blue color refers to the gamma rays filtered out due to their energy/angle. 

The average energy of the gamma rays produced is $\sim$15~MeV. However, due to the limited geometrical aperture, the gamma rays below 15 MeV do not reach the detector. The average energy of the gamma rays reaching the detector is therefore 24~MeV. Later, these results are used for the Geant4 simulation and calibration of the calorimeter.

\section{Measurements}
\label{sec:measur}
\subsection{Scanning procedure and first data}
The FPC was commissioned in October 2010 and Compton collisions were recorded on the first attempt (26 October). Before data taking we investigated how the intensity of the signal is affected by the position of the laser beam waist with respect to the electron beam position and synchronization between the laser and electron beam by doing the position and the phase scans.

\begin{figure}[htbp]
\begin{center}
\includegraphics[width=0.8\textwidth]{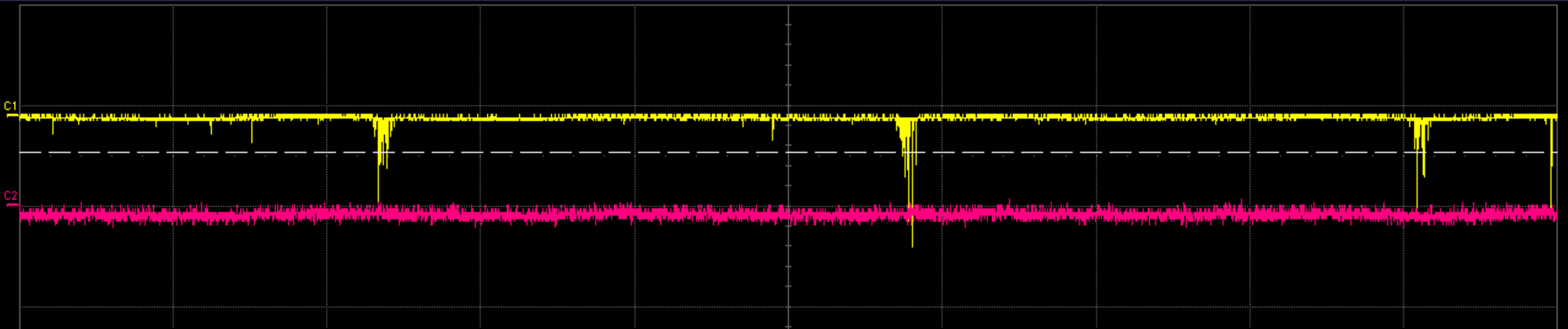}
\caption{Example of signal observed on our oscilloscope while scanning the parameters space to find the collision area. The yellow line is the signal from the PMT. The duration of the waveform is  10~ms and the vertical scale is 200~mV (on this day the signal was amplified).}
\label{fig:signal_bursts}
\end{center}
\end{figure}

The search for the collisions area between the electrons and the laser photons has several degrees of freedom and is not straightforward. The two most sensitive degrees of freedom are the vertical position of the FPC (dimension orthogonal to the plane of the beams) and the relative phase between the laser and the electrons (time dimension). The search for collisions is performed in several steps:

\begin{itemize}
\item  The FPC frequency is offset by at least 100~Hz with respect to the ATF frequency. By doing this the relative phase between the electrons and the laser is automatically scanned at a rate of at least 100~Hz.
\item The vertical position of the FPC is swept slowly.
\item  While performing this vertical scan we look on the data acquisition oscilloscope for the apparition of short bursts in the output of the PMT (see figure~\ref{fig:signal_bursts}). The length and the frequency of these bursts are related to the difference between the frequency of the cavity and the frequency of the ATF clock.
\item Once these bursts have been found the vertical position of the laser is adjusted to maximize their intensity.
\item  At this stage the laser frequency can be locked on the ATF frequency.
\item The phase of the laser with respect to the electron beam must be adjusted to optimize the intensity of the signal observed on the PMT output\footnote{Our experience shows that the variation of this phase from day to day is small however we do observe small drifts which require a new phase scan on each run.}. 
\end{itemize}

The bursts correspond to the time at which the phase of the laser matches that of the electrons. The distances between the bursts corresponds to the difference between the ATF frequency and the laser frequency. Each burst contains several peaks spaced by the duration of two DR revolution. The red line records the injection trigger (the trigger itself is off the screen). It was checked that the bursts on the yellow lines disappear when the cavity is moved vertically by a distance greater than the beam size.

\subsection{Example of data}

\begin{figure}[htb]
   \centering
   \includegraphics[width=0.5\textwidth]{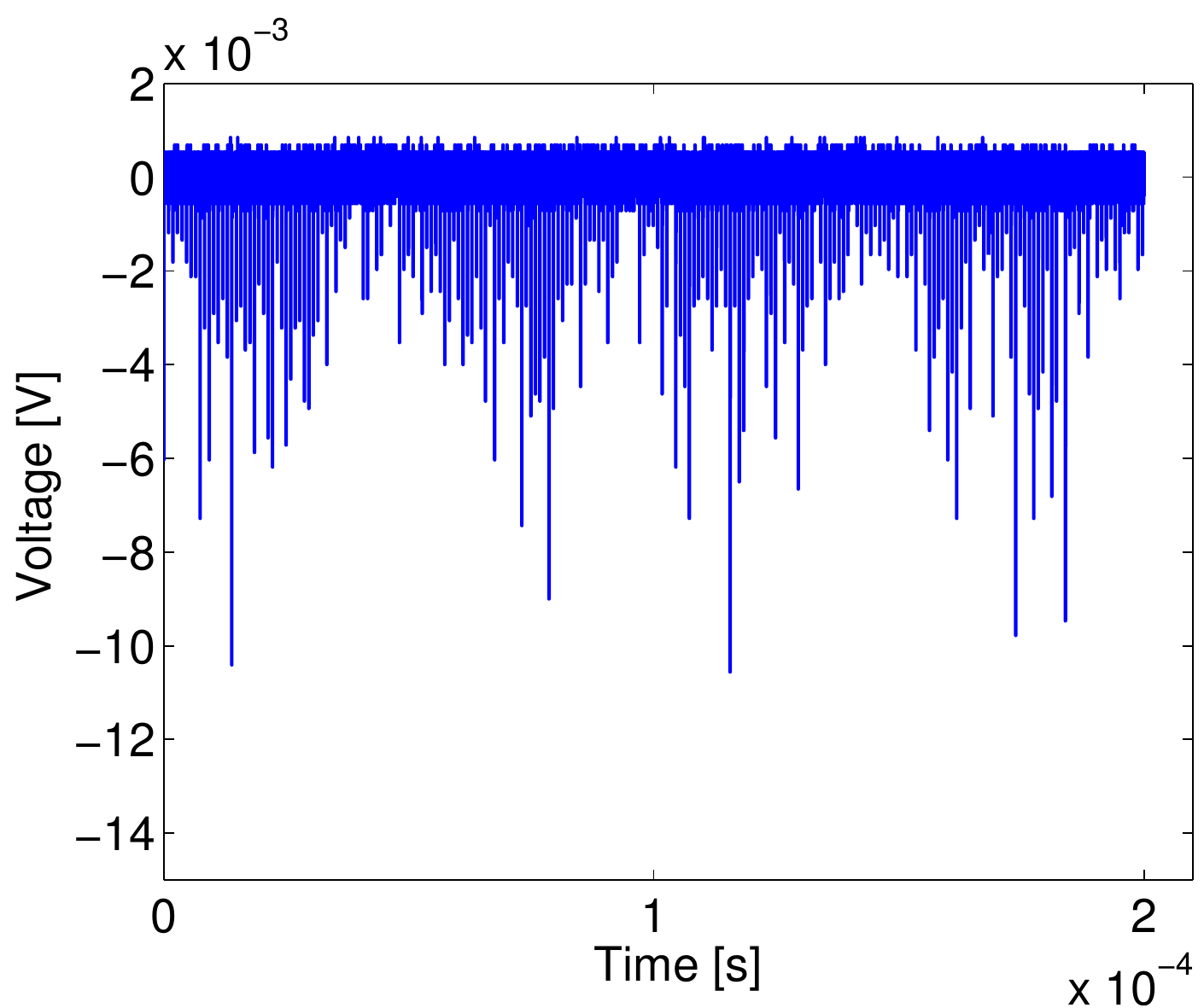}
   \caption{The typical signal shape for the high energy gamma rays produced by Compton scattering observed at the anode output of the PMT. Each spike on the picture corresponds to the gamma production after successive bunch crossings over 0.2~ms. Data shown correspond to a single bunch stored in the ATF DR.}
   \label{fig:typicalWF}
\end{figure}

During data taking we record the signal from the PMT as well as the 357 MHz ATF clock and the laser power transmitted by the FPC measured by a photodiode. 
To avoid the mistaking noise from the injection or extraction kickers a typical data acquisition starts at least 200~ms after the injection trigger is received. At this time we expect the beam to be almost fully damped. 
A typical signal waveform from the calorimeter can be observed on figure~\ref{fig:typicalWF}. A full waveform contains approximately 200 000 samples spaced by  1~ns.  

\subsection{Data analysis} 
The raw data recorded contain the intensity of the PMT output as a function of time (see figure~\ref{fig:typicalWF}). Data analysis was carried out to extract the intensity of the Compton signal and remove unwanted backgrounds. 

\begin{figure}[htbp]
  \centering
  \subfloat[]{\label{fig:stacking}\includegraphics[width=0.47\textwidth]{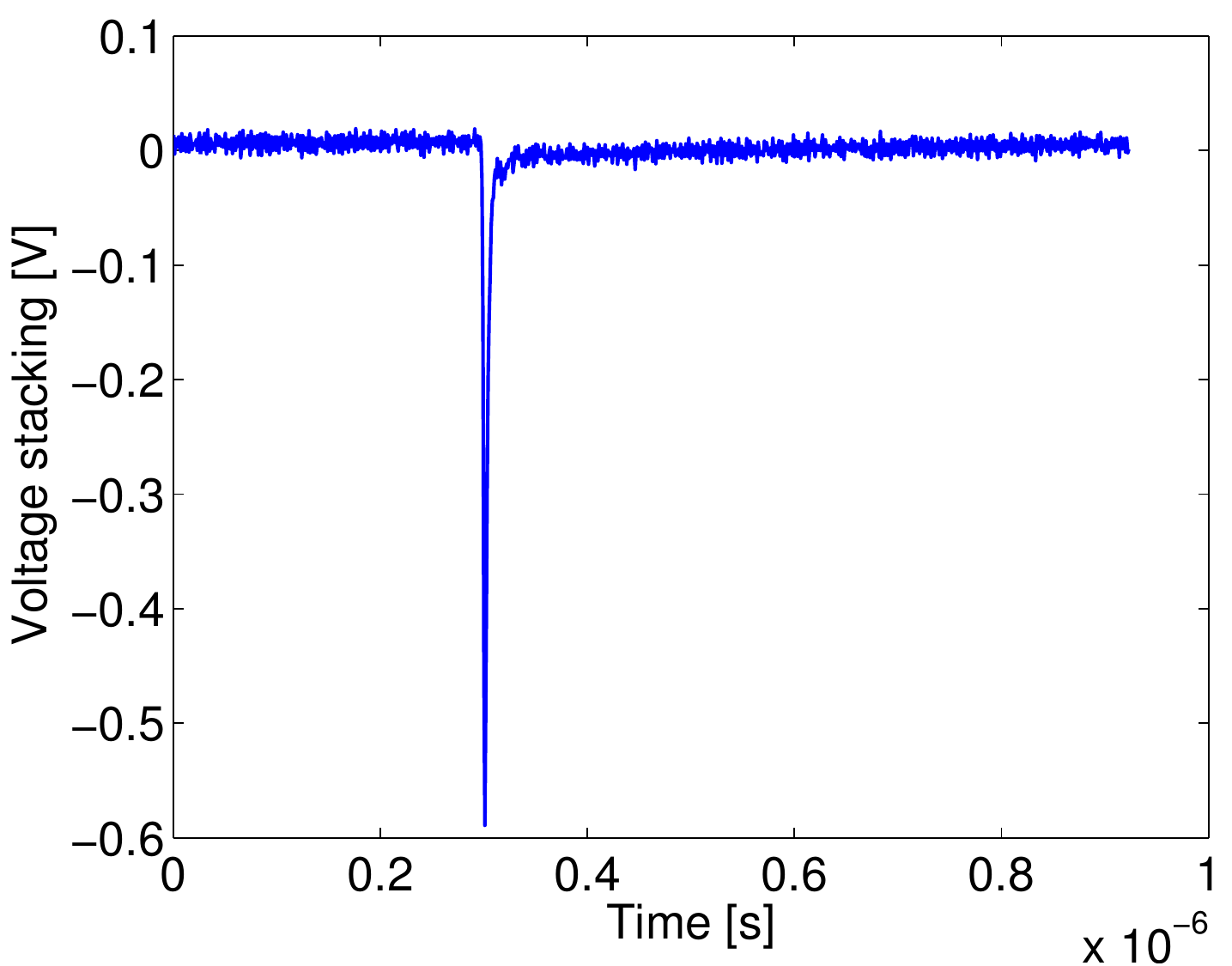}}    
  \hspace{5mm}            
  \subfloat[]{\label{fig:gates}\includegraphics[width=0.47\textwidth]{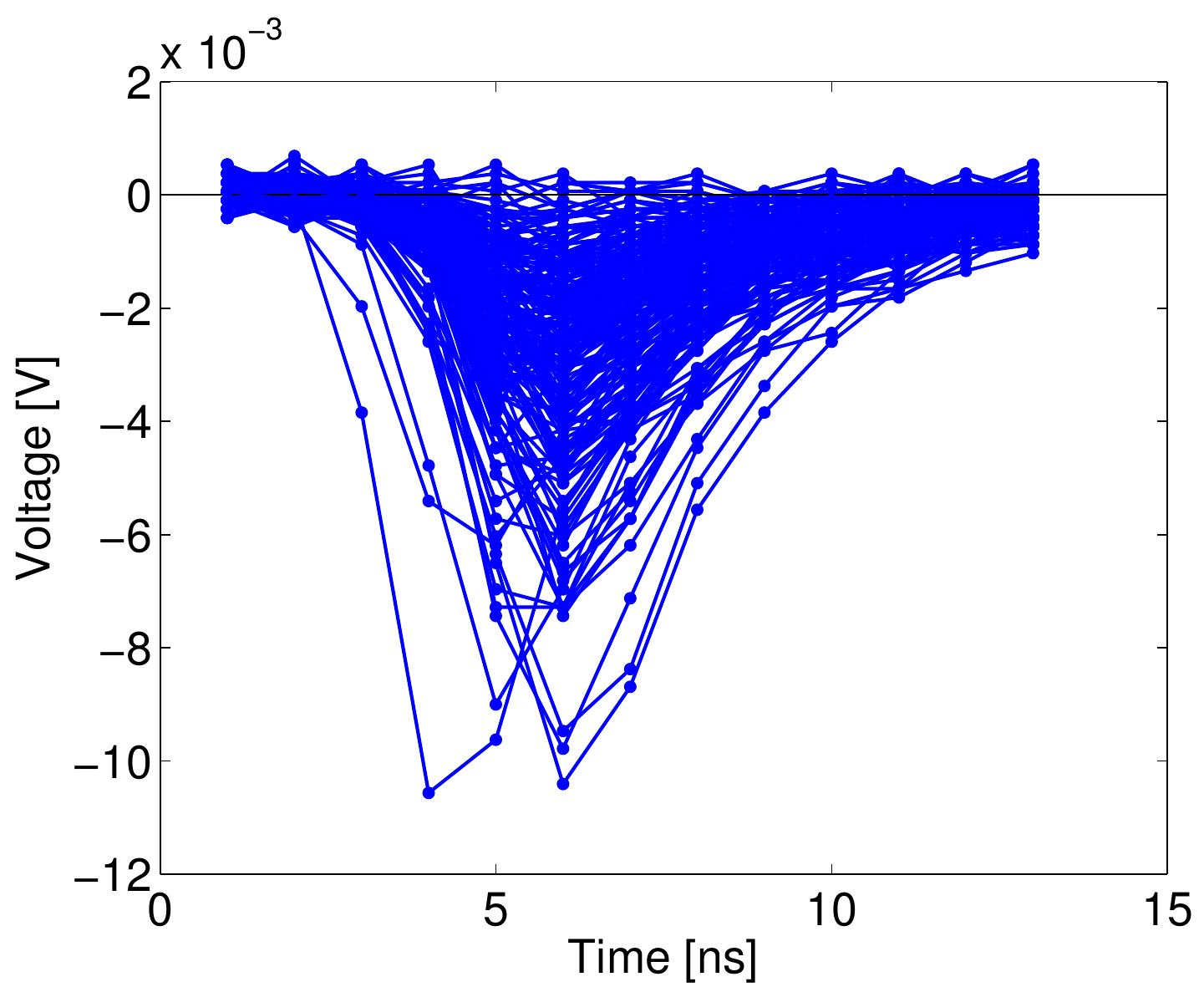}}
  \caption{Stacking of all the 924~ns periods from the data file presented on figure~\protect\ref{fig:typicalWF} (figure~\protect\ref{fig:stacking}). This stacking is used to find the number of the peaks and their timing. Time profile of the signals for all the periods from the same file is shown on figure~\protect\ref{fig:gates}.}
  \label{fig:stack_gates}
\end{figure}

As the data were acquired during the different ATF runs with different filling mode of the DR (1~train,  2~trains, 3~trains stored in the ATF DR) the correct number of Compton peaks and their timing must be found. The 357~MHz ATF clock is used to define the beginning of a 924~ns periods corresponding to the occurrence of the Compton signal (two ATF DR revolutions). 
All 924~ns periods belonging to one data file (usually 0.2~ms long) are superimposed on top of each other to find the number of the peaks per revolution and their time of arrival with respect to the beginning of the period. Such technique is used to 
enhance the signal over the background reducing at the same time the electronic noise. This results in a more precise way for the estimation of the number of the peaks and their timing. An example of  stacking can be seen on figure~\ref{fig:stacking}.
 
Once the number of peaks and their positions within the period is found we define a gate around the position where the Compton signal is expected. We use this gate to calculate the height and integral of the peak. The Compton peak height and its integral are our measure of the energy deposited by the gamma rays in the calorimeter. The length of the gate is set to 12~ns in order to entirely contain the signal thus ensuring a correct evaluation  of the energy deposited. The distribution of such gates around the peak is presented on figure~\ref{fig:gates}.

\begin{figure}[htbp]
\centering
\subfloat[]{\label{fig:PH}\includegraphics[width=0.47\textwidth]{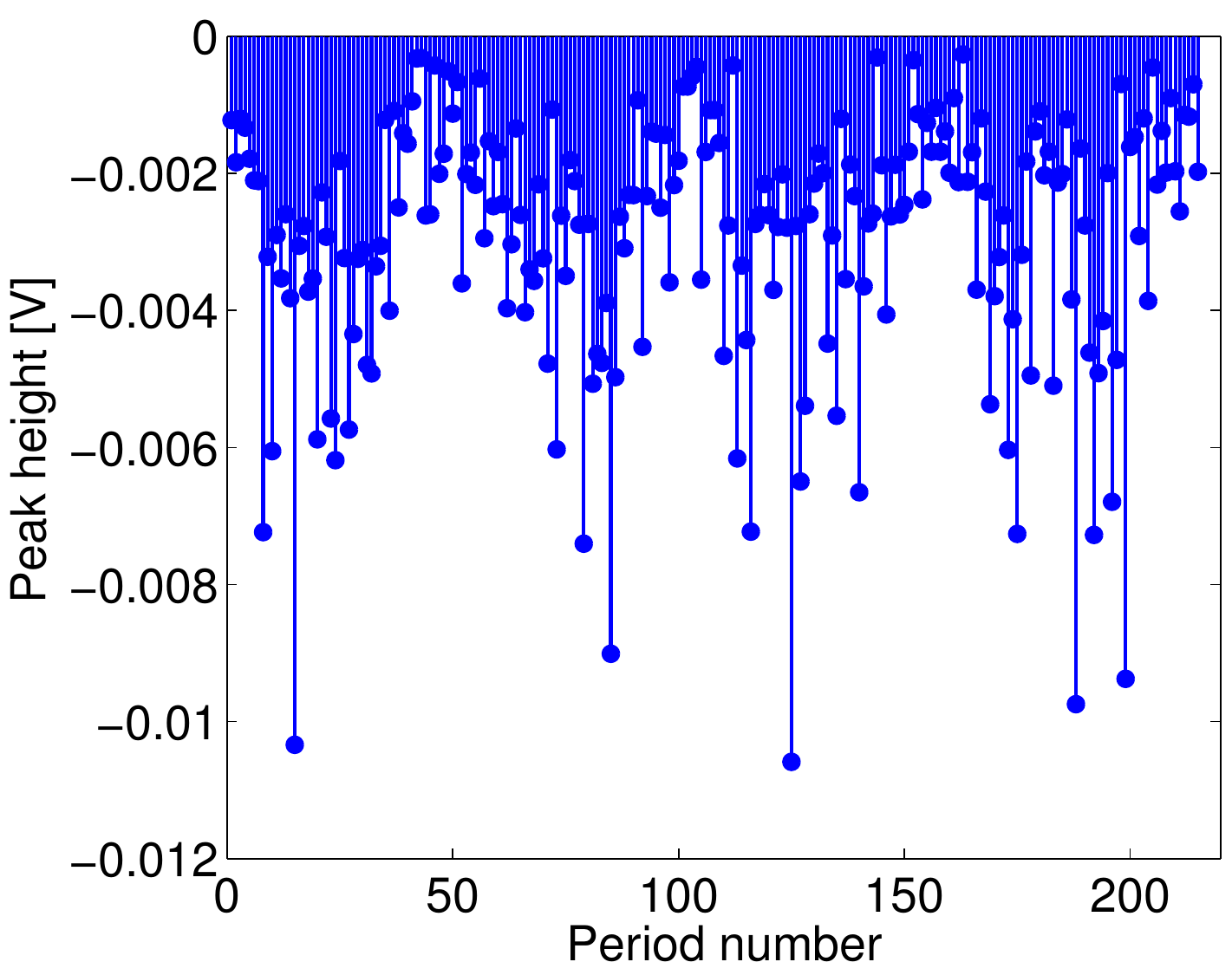}}    
\hspace{5mm}            
\subfloat[]{\label{fig:slope}\includegraphics[width=0.47\textwidth]{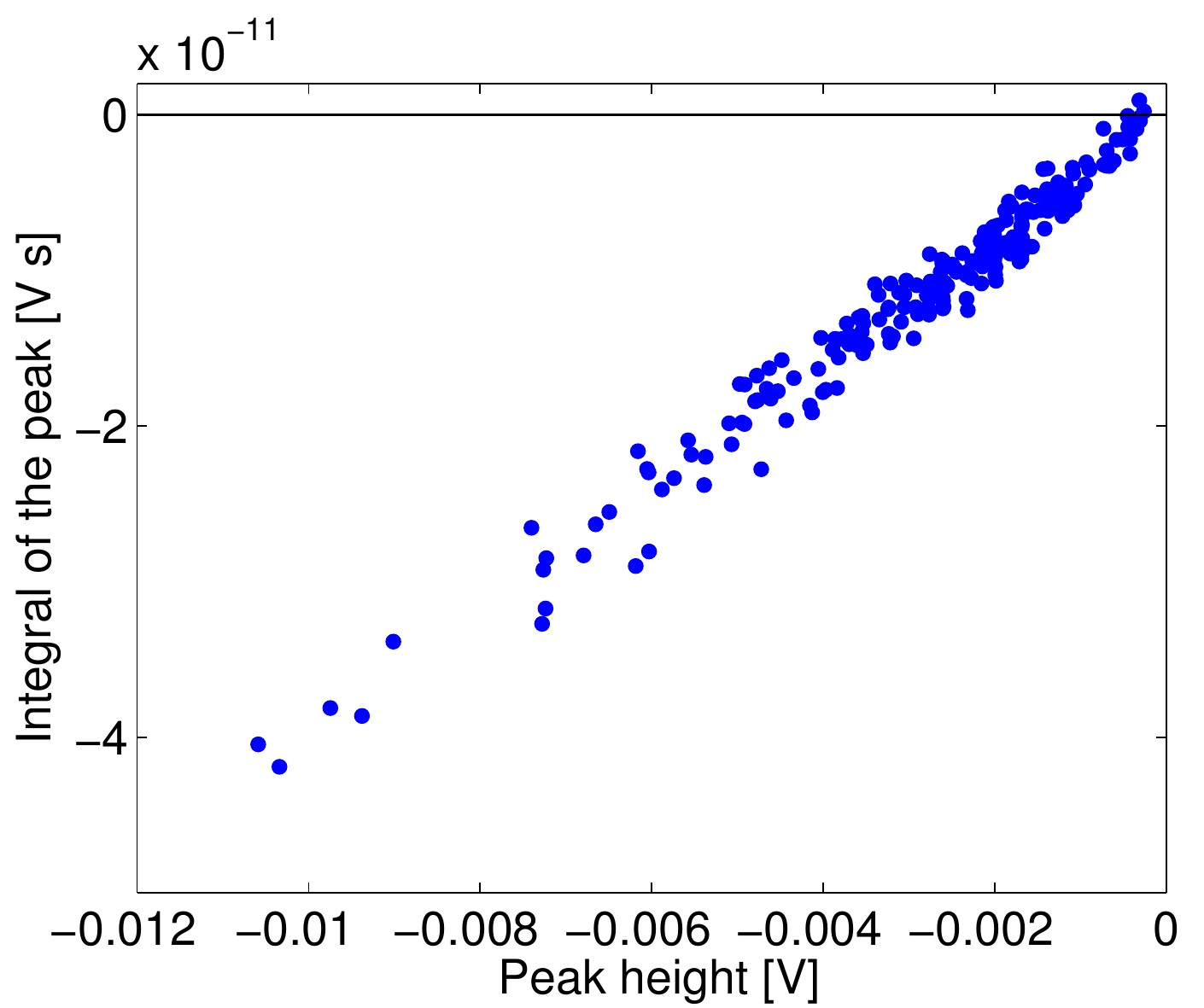}}
\caption{Figure~\protect\ref{fig:PH} gives an example of the peak height distribution from the data file presented on figure~\protect\ref{fig:typicalWF}. Peak height vs. peak integral for the same data file is shown on figure~\protect\ref{fig:slope}.}
\label{fig:PHSlope}
\end{figure}

The background level and its RMS are calculated within each period and are subtracted from the corresponding peak height and peak integral. The background level is defined as the mean over the timebase corresponding to a given period  excluding the gate containing the signal. The average background level and its RMS for our data sample is estimated to be around 7~$\mu$V and 0.2~mV respectively. Finally, the peak height and peak integral are calculated for every peak within the period (see figure~\ref{fig:PH}). 

The shape of the detector's response creates a linear relation between the total charge and the maximum charge measured. This can be seen as a correlation between the calculated peak height and peak integral as shown on figure~\ref{fig:slope}

\begin{figure}[htbp]
  \centering
  \subfloat[]{\label{fig:PP}\includegraphics[width=0.33\textwidth]{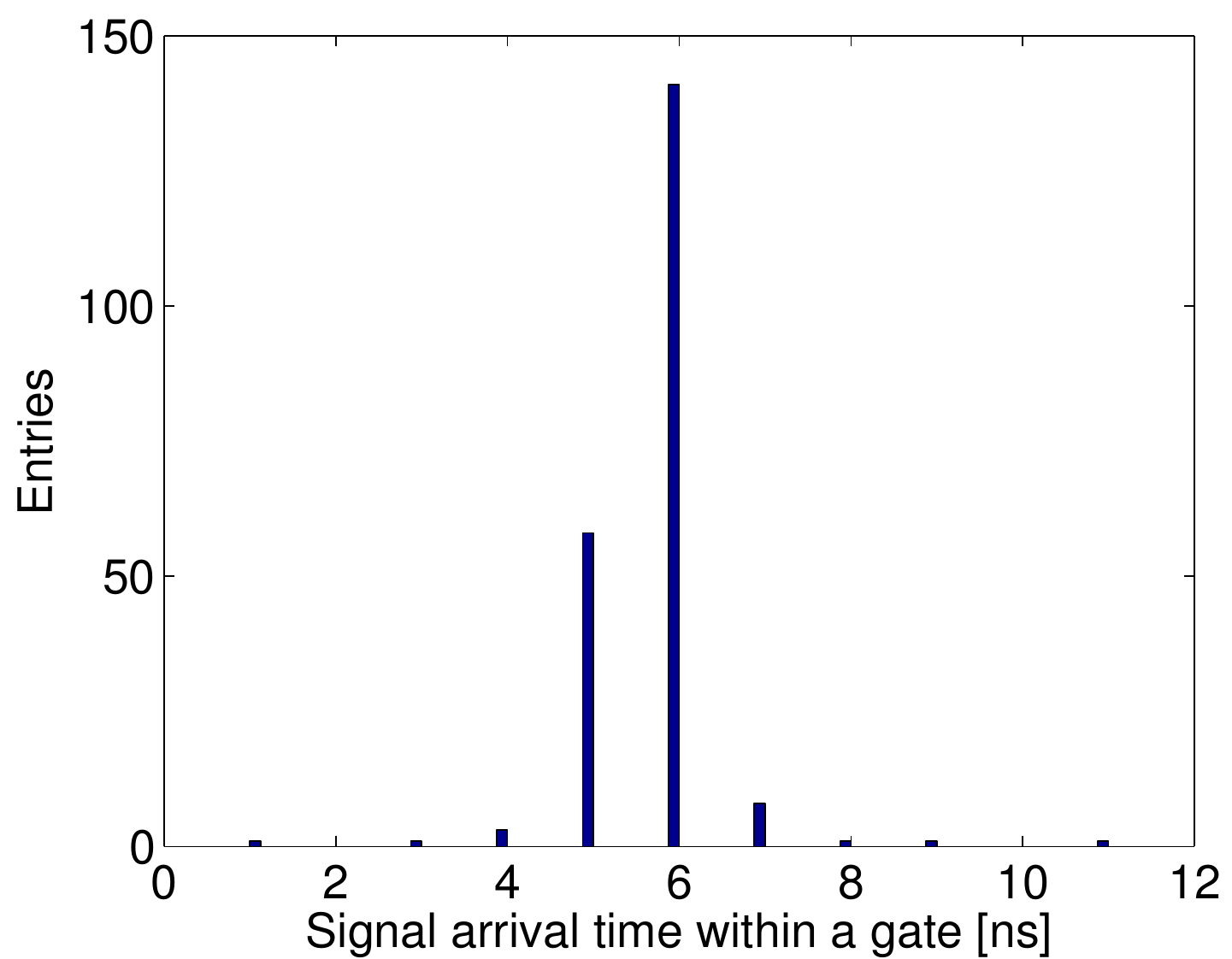}}    
  \subfloat[]{\label{fig:ELPshape}\includegraphics[width=0.33\textwidth]{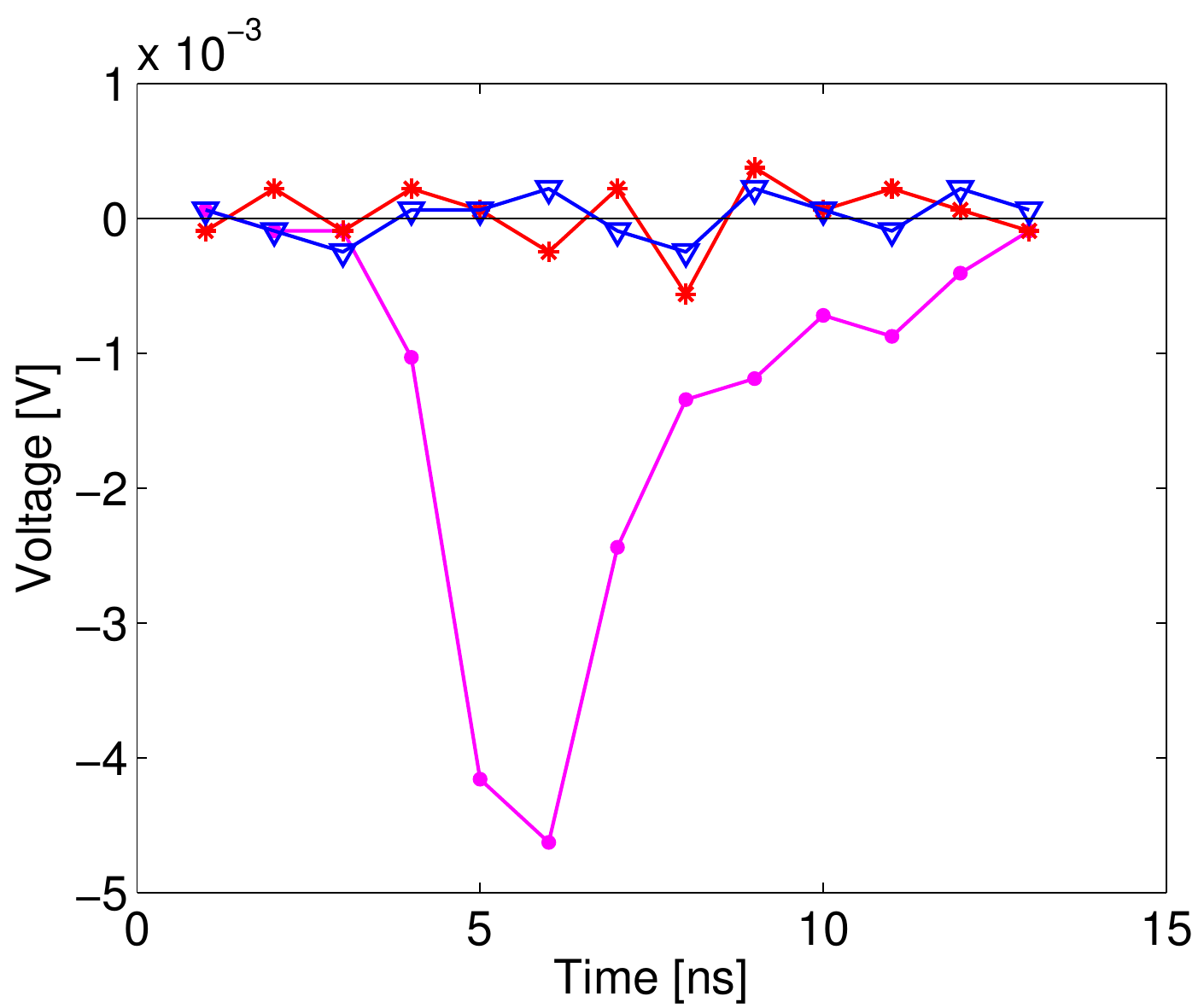}}
  \subfloat[]{\label{fig:Qslope}\includegraphics[width=0.33\textwidth]{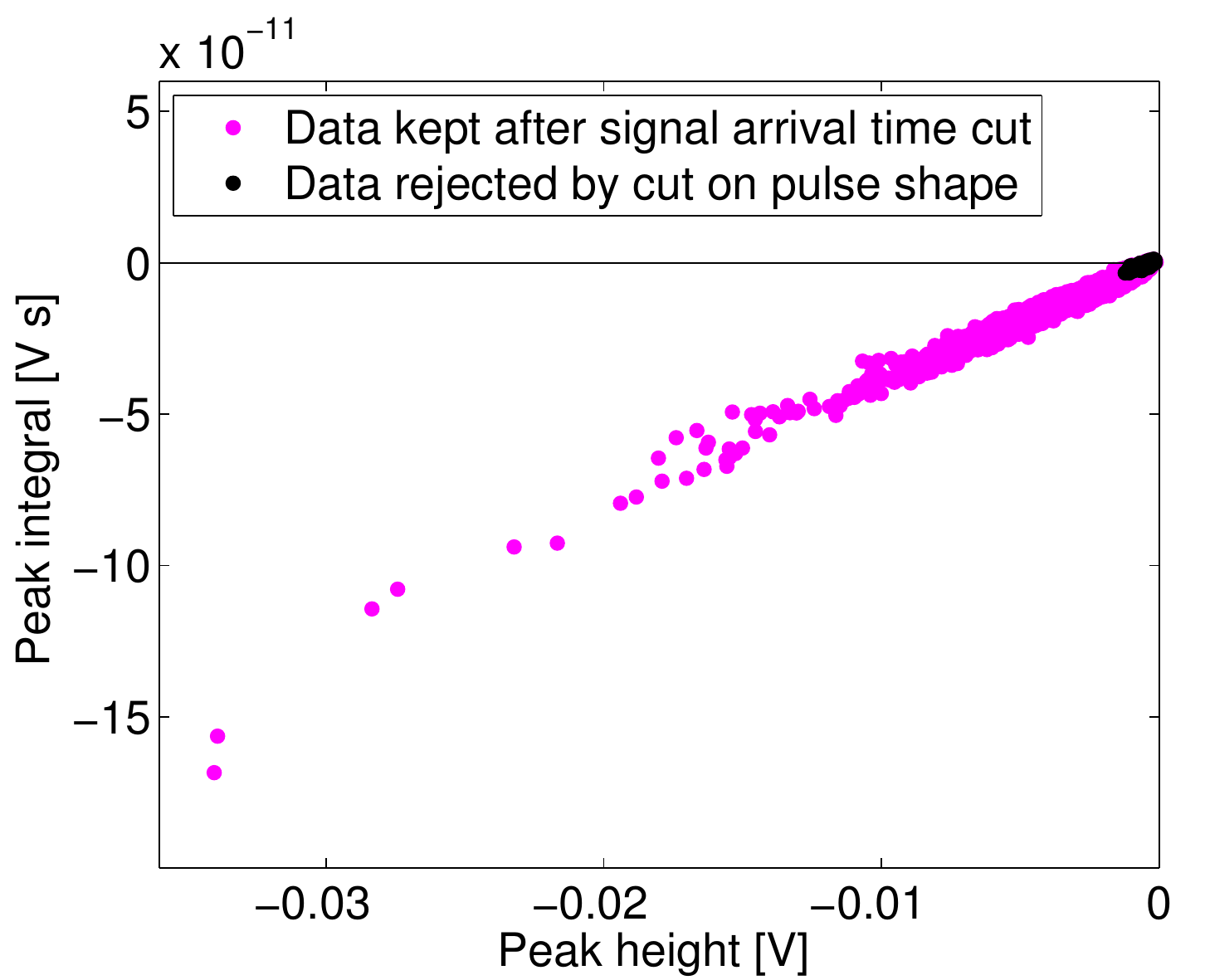}}
  \caption{Quality cut illustration (figures~\protect\ref{fig:PP} and \protect\ref{fig:ELPshape}) shown together with their effect on all the data sample (figure~\protect\ref{fig:Qslope}). Figure~\protect\ref{fig:PP}:  the distribution of the signal arrival time within the gate for all the signals found in the data file presented on figure~\protect\ref{fig:typicalWF}. Figure~\protect\ref{fig:ELPshape} gives an example of a correct signal (magenta line) as opposed to the noise signal (blue, red line). Figure~\protect\ref{fig:Qslope} shows the data sample remaining after the cut on the signal arrival time applied (magenta points)  together with the events which are rejected due to the cut on the shape of the pulse (data points tagged in black).}
  \label{fig:CUT}
\end{figure}

\paragraph{Quality of the data.}
Different quality cuts are applied to restrict the analysis to a high purity sample. By putting limitations on  them we can reject noise and obtain a set of good quality data (see figure~\ref{fig:CUT}). 


We want all the Compton peaks within one file to have the same phase with respect to the ATF clock. It22 is achieved, in this case, by requiring the Compton signal to arrive 6~ns after the beginning of the gate (see figure~\ref{fig:PP}).

Sometimes, the noise can dominate the signal (see figure~\ref{fig:ELPshape}). Quite often, these signals can even lead to an integral value with the wrong sign. This is illustrated on figure~\ref{fig:slope} by the dots with positive values of the integral of the peak. Such events have to be considered as picked up noise and must be rejected. 
For this, we introduce a variable:
\begin{equation}
 V=\frac{2 I_{max}(t_{0})}{ I(t_{0}-1) + I(t_{0}+1)},
\end{equation}
where $ I_{max}(t_{0})$ is the intensity of the signal taken at the time when it reaches maximum, $I(t_{0}-1)$ and $I(t_{0}+1)$ are the intensities of the signal in the two nearest
data points.
By imposing $1\leq V \leq10$ the noise is filtered out as $V$ reaches either negative or very high positive values for noise signal. 
 On figure~\ref{fig:ELPshape} the magenta line corresponds to the real Compton signal for which our estimator variable  $V$ is about 1 while the red and blue lines are for the noise events and in this case $V$ equals to -2 and 17 respectively. 

The effect of the cuts mentioned above is shown on figure~\ref{fig:Qslope}. Cut on the signal arrival time preserve good linear relation between the peak height and peak integral apart from a few high intensity data points for which our data acquisition system saturated. The cut on the shape of the pulse helps to remove noise. However, some low intensity events still remain which correspond to the low energy deposited by the gamma rays in the calorimeter.

Approximately half of the data taken and presented in this study have been rejected after the different quality cuts have been applied. All the results commented below are based on high quality data sample.
    
\paragraph{Background} 
While analysing the data we noticed that some data files contain peaks spaced by about 462~ns corresponding to one ATF DR revolution period.  With the experimental setup we use this is not possible because the electron bunch can not interact with laser pulses on two consecutive turns (see section~\ref{expsetup}). Such turn-by-turn background could be caused by the electrons hitting something inside the accelerator. To enforce high purity data, the files with such background are rejected.

\begin{figure}[htbp]
\begin{center}
\subfloat[]{\label{fig:LP_histo}\includegraphics[width=0.47\textwidth]{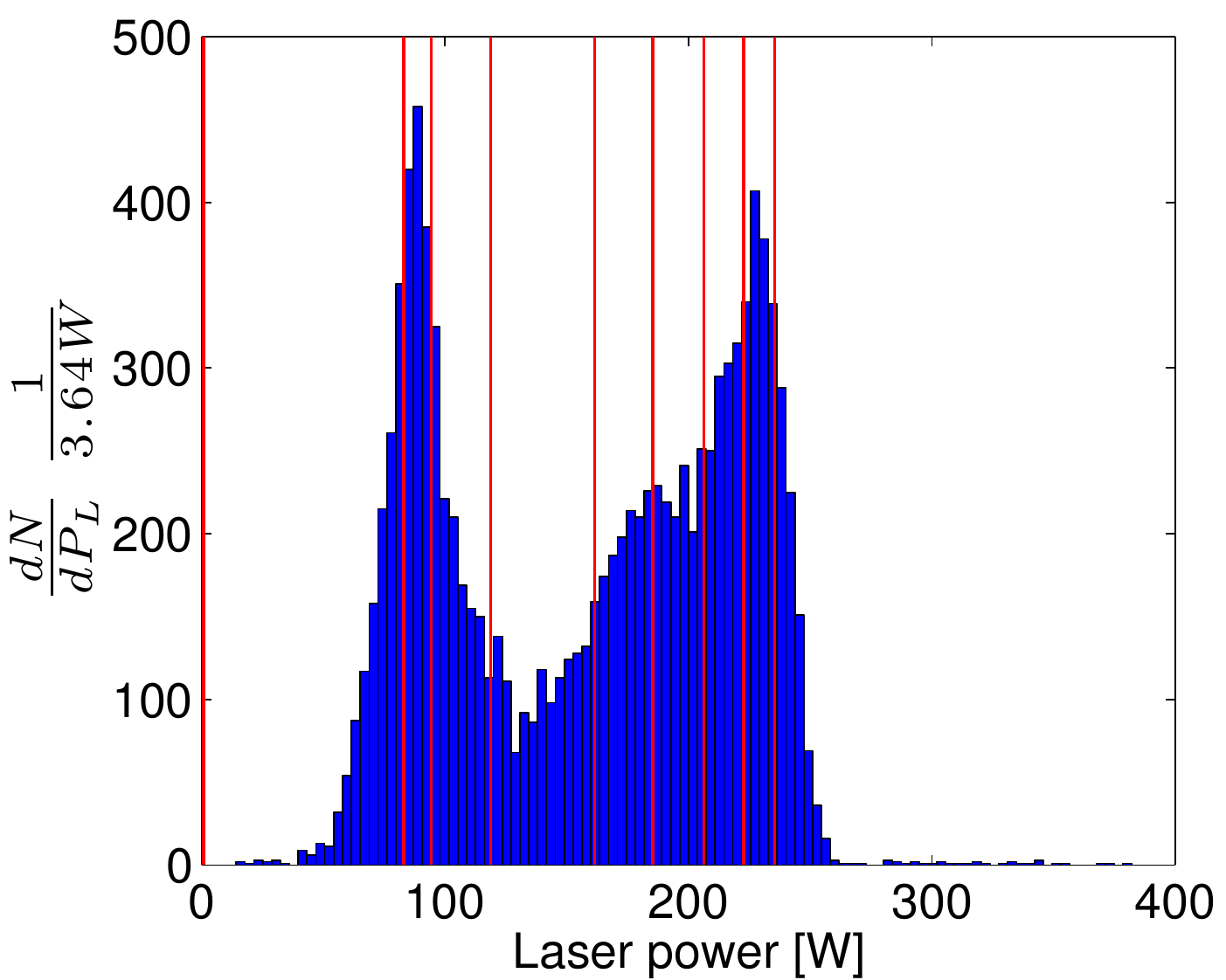}}
\hspace{3mm}            
\subfloat[]{\label{fig:PH_LP}\includegraphics[width=0.49\textwidth]{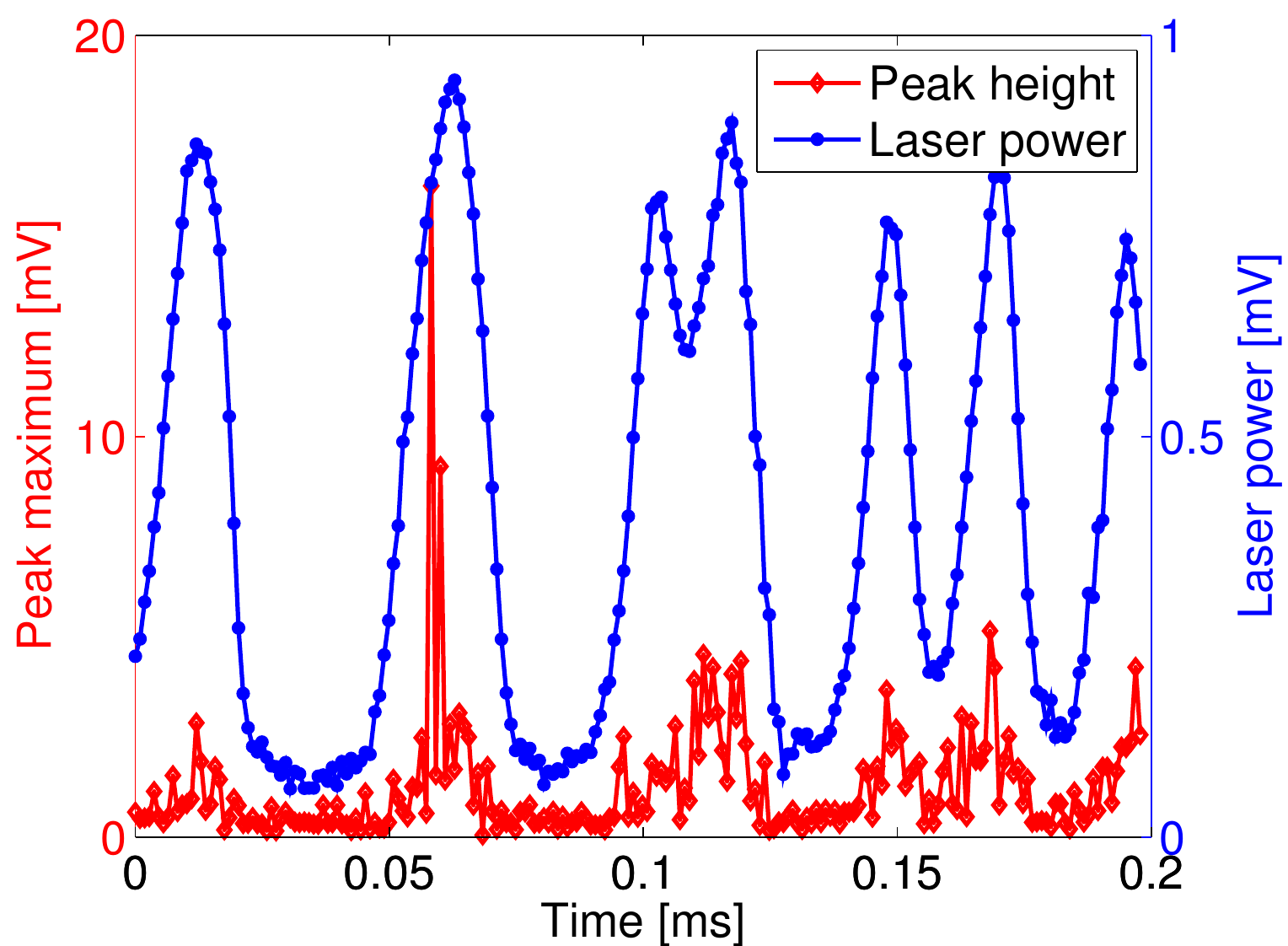}}
\caption{Figure~\protect\ref{fig:LP_histo} shows the histogram of the laser power stored in the FPC normalized by its maximum value. Red vertical lines divide the distribution into the nine laser power bins of approximately the same population each (see section~\protect\ref{sec:results}). Figure~\protect\ref{fig:PH_LP} gives an example of the correlation between the transmitted FPC laser power and the gamma ray production intensity for a given data file. This can be seen by the correspondence between the peak height distribution (blue line) and the laser power measured by the photodiode at the same time (red line). }
\label{fig:laser_power}
\end{center}
\end{figure}

\paragraph{Laser power}  \label{par:LP}
On the day where the data presented in this article were taken, the average power stored in the FPC was approximately in average 160~W. It was measured by a photodiode placed behind   one of the mirrors of the FPC and calibrated with a power-meter. Figure~\ref{fig:LP_histo} shows the distribution of stored FPC laser power measured by the photodiode. As one can see, the FPC laser power  experienced significant fluctuations which can be explained by several resonances in the FPC feedback system. 
Note that the feedback loop used to stabilize the laser power stored in the FPC had not yet been optimized at the time when these data were taken.
On figure~\ref{fig:PH_LP} the strong correlation between the laser power and the peak height  distribution is clearly observed.

\subsection{Calibration}

In the energy range (tens of MeV) of interest, the calorimeter can be calibrated by using cosmic rays muons. This calibration gives us  a relation between the voltage measured on the oscilloscope and the energy deposited in the calorimeter. 

\begin{figure}[htbp]
\centering
\subfloat[]{\label{fig:calib1}\includegraphics[width=0.47\textwidth]{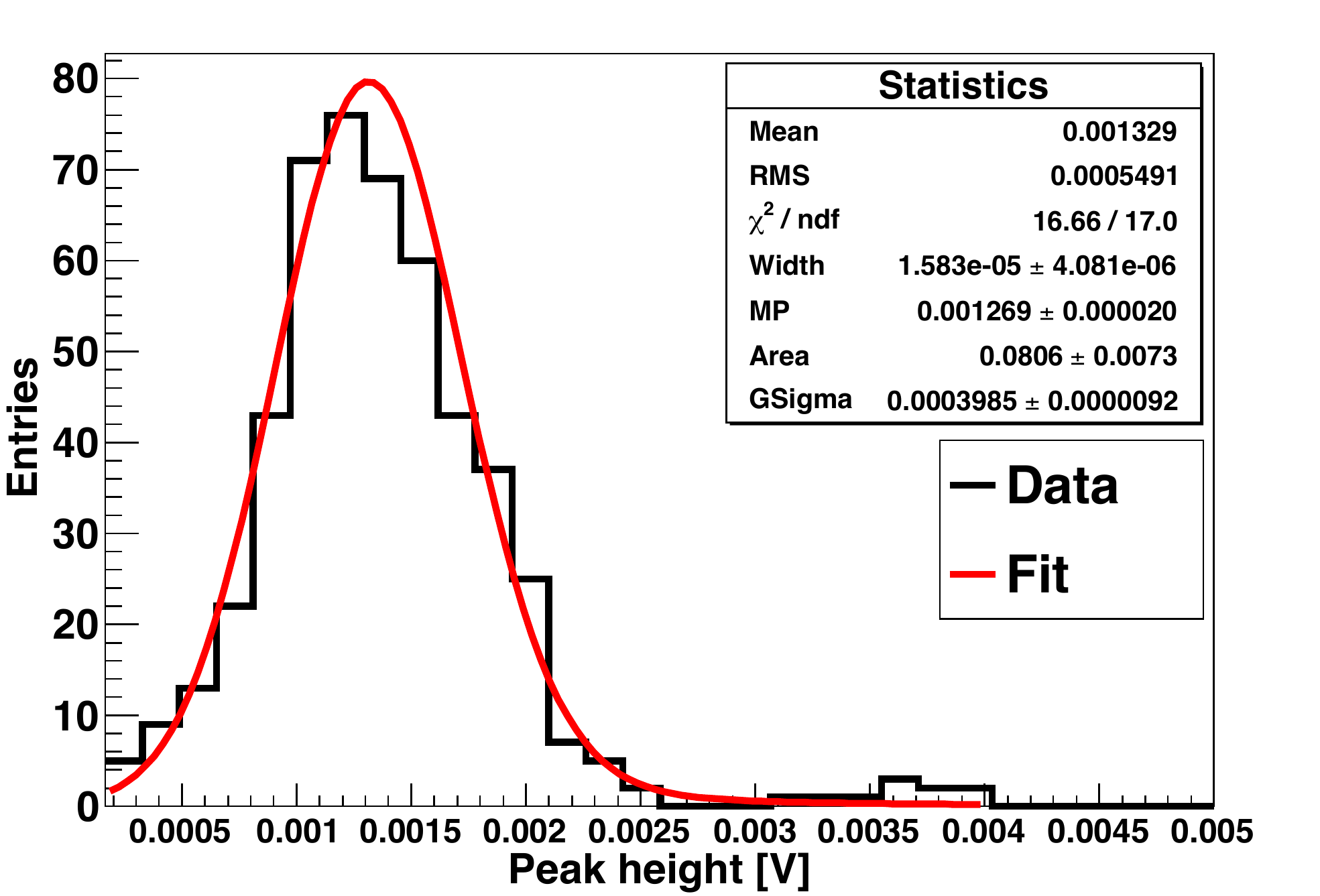}}    
\hspace{5mm}            
\subfloat[]{\label{fig:calib2}\includegraphics[width=0.49\textwidth]{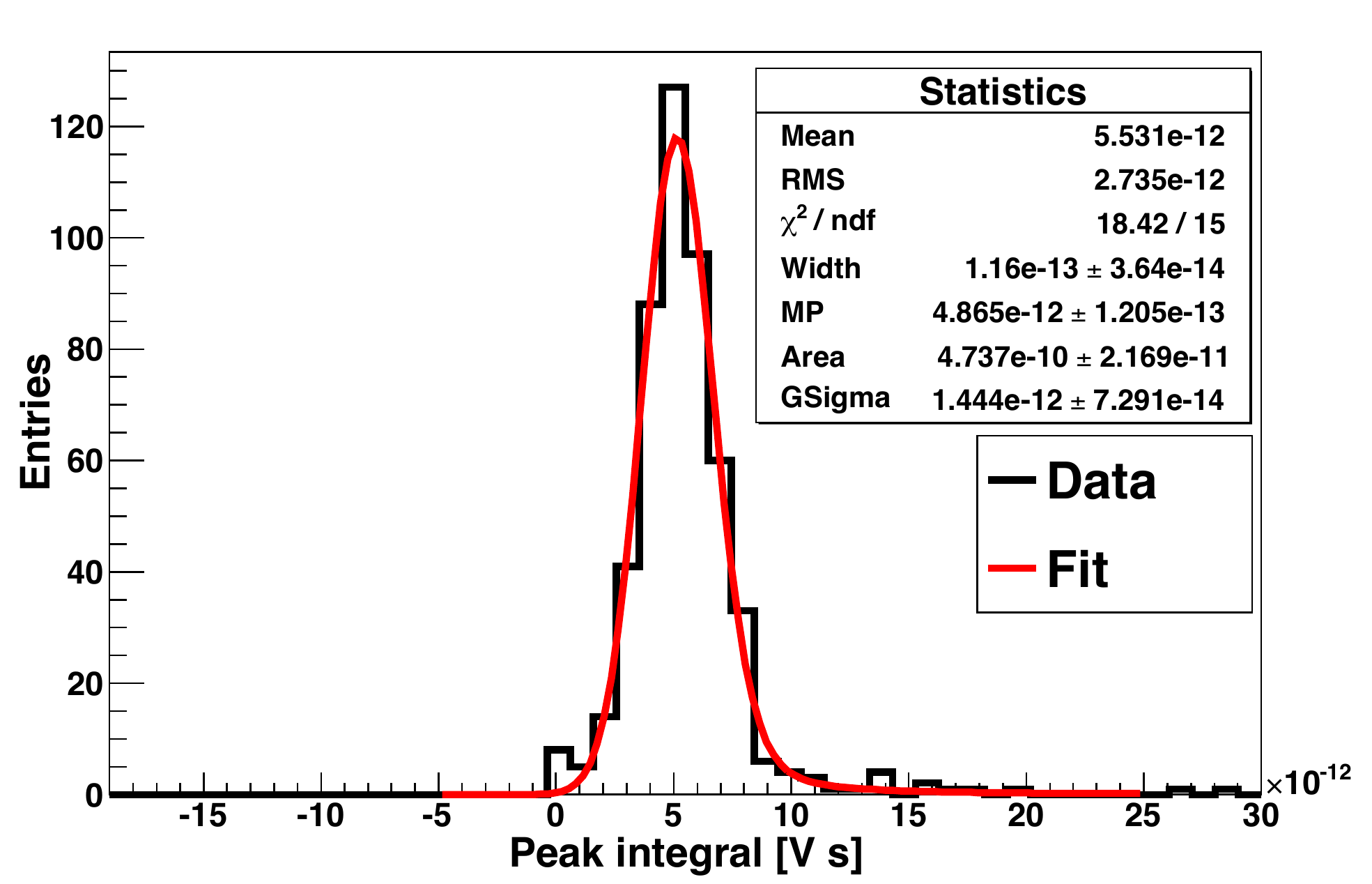}}
\caption{Energy loss distribution of the cosmic muons in the calorimeter. Figure~\protect\ref{fig:calib1} shows the histogram of the peak heights and figure~\protect\ref{fig:calib2}  shows the histogram of the peak integrals. A Landau distribution fit with parameters is shown in both cases.}
\label{fig:calib}
\end{figure}

In calibration mode the coincidences between two plastic scintillators, one placed at the top and the other at the bottom of the calorimeter, are used to trigger the data acquisition. The signal from the PMT is then digitized by the oscilloscope used to acquire the Compton data. Calibration measurements were performed before each data taking run. 
A typical measured distribution of pulse heights and pulse integrals is shown on figure~\ref{fig:calib}.

For detectors of moderate thickness such as the one we are using, the energy loss probability distribution is described by the Landau distribution~\cite{landau1944energy}. The most probable energy loss of cosmic muons is the most probable value (MPV) of the Landau distribution.

According to our calibration, the MPV for the integrated signal is (4.87~$\pm$~0.12)~mV$\cdot$ns and the MPV for the peak height distribution is (1.27~$\pm$~0.02)~mV. Assuming minimum ionization, the energy deposition in the calorimeter is 6.374~MeV/cm ~\cite{groom2011atomic}. This value corresponds to approximately 45~MeV of energy deposited by a muon passing through the calorimeter. However, this value should be corrected to account for the geometry and the acceptance of the detection system. 

To do so, the measure should be scaled by the ratio between the total number of optical photons produced in the  $BaF_2$ crystal and those optical photons which are within the PMT acceptance. Another correction factor appears due to the fact that the two surrounding plastic scintillators are not fully covering the surface of the $BaF_2$ crystal. Therefore, the previous factor has to be corrected to take into account this difference in overlapping surfaces. To address these two issues we used a Geant4 simulation of the detector which gave a correction factor of 0.745~$\pm$~0.001(stat)~$\pm$~0.04(syst).

Once, all corrections have been applied, a peak height of 1 mV and peak integral of 1~mV$\cdot$~ns is equivalent in average to (26.6~$\pm$~1.9)~MeV and (6.9~$\pm$~0.5)~MeV of energy deposited in the calorimeter respectively.

\section{Results}
\label{sec:results}
The data analysis performed in this study were done to evaluate the number of gamma rays produced during the collisions between the electrons and laser photons.

\begin{figure}[htbp]
\centering
\subfloat[]{\label{fig:HQ_PI}\includegraphics[width=0.47\textwidth]{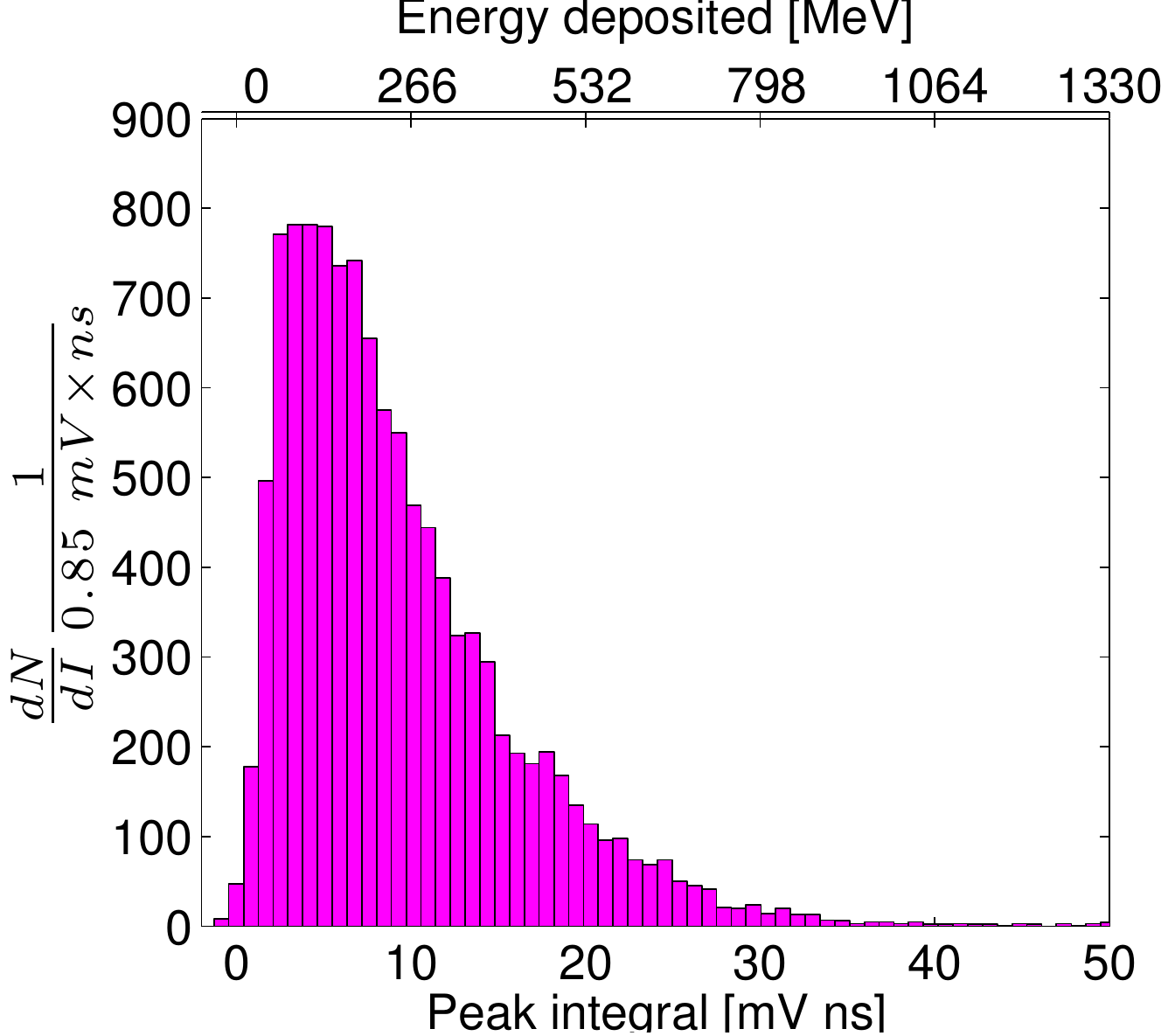}}    
\hspace{5mm}            
\subfloat[]{\label{fig:cHQ_PI_norm}\includegraphics[width=0.47\textwidth]{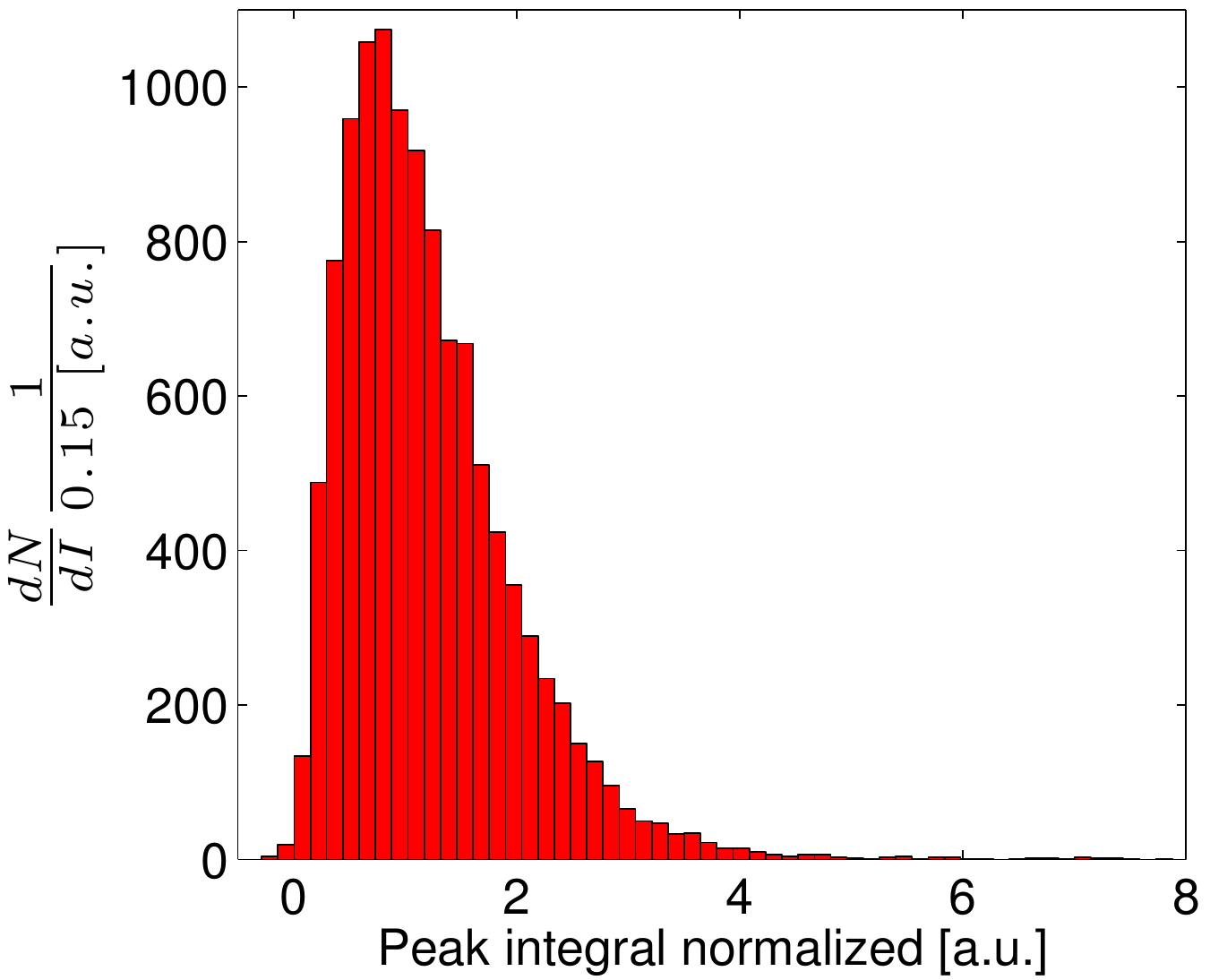}}
\caption{Spectrum of the gamma rays produced for all laser powers. Figure~\protect\ref{fig:HQ_PI}  shows the distribution of the integrals of the Compton peaks for the high quality data sample whereas the figure~\protect\ref{fig:cHQ_PI_norm} shows the same spectrum normalized by the laser power. }
\label{fig:spectra}
\end{figure}

The spectrum of the gamma rays is shown on figure~\ref{fig:HQ_PI} and~\ref{fig:cHQ_PI_norm}. It represents the distribution of the energy deposited in the calorimeter expressed by the peak integrals\footnote{The similar spectra could be obtained by using the Compton peak height as a measure of energy deposited in the calorimeter.}. 

Using our calibration, we can estimate the average number of scattered gammas per bunch crossing. As it was shown in section~\ref{sec:expectgflux}, the mean energy of scattered gammas we measure is 24~MeV. Since the average energy deposited per bunch crossing is 65.1~$\pm$~4.9~MeV, we deduce that approximately 2.7~$\pm$~0.2~gammas are produced in average per bunch crossing (for an average laser power stored in the FPC of about 160 W). This rate of about 2.7~gammas per bunch crossing was successfully sustained over 6~hours until the end of our run.
For the given collision repetition frequency in the ATF DR of about 1 MHz, the flux of gamma rays achieved is about $3\times10^6$~gammas per second  (this does not take into account the 0.75 duty cycle of the ATF).

\begin{figure}[htbp]
\centering
\subfloat[]{\label{fig:mean_spectrum_PS}\includegraphics[width=0.47\textwidth]{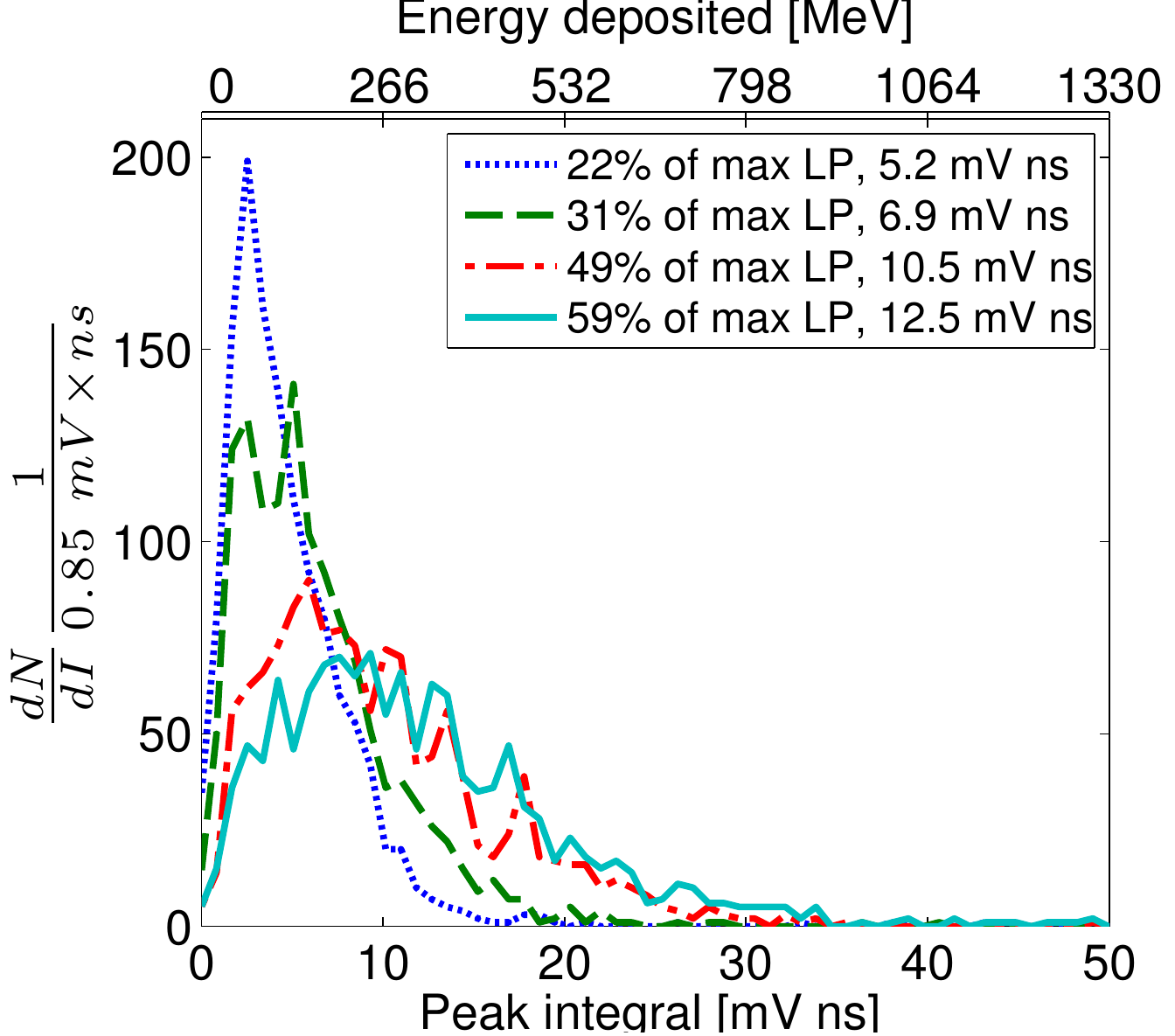}}    
\hspace{5mm}            
\subfloat[]{\label{fig:spectrum_PS}\includegraphics[width=0.47\textwidth]{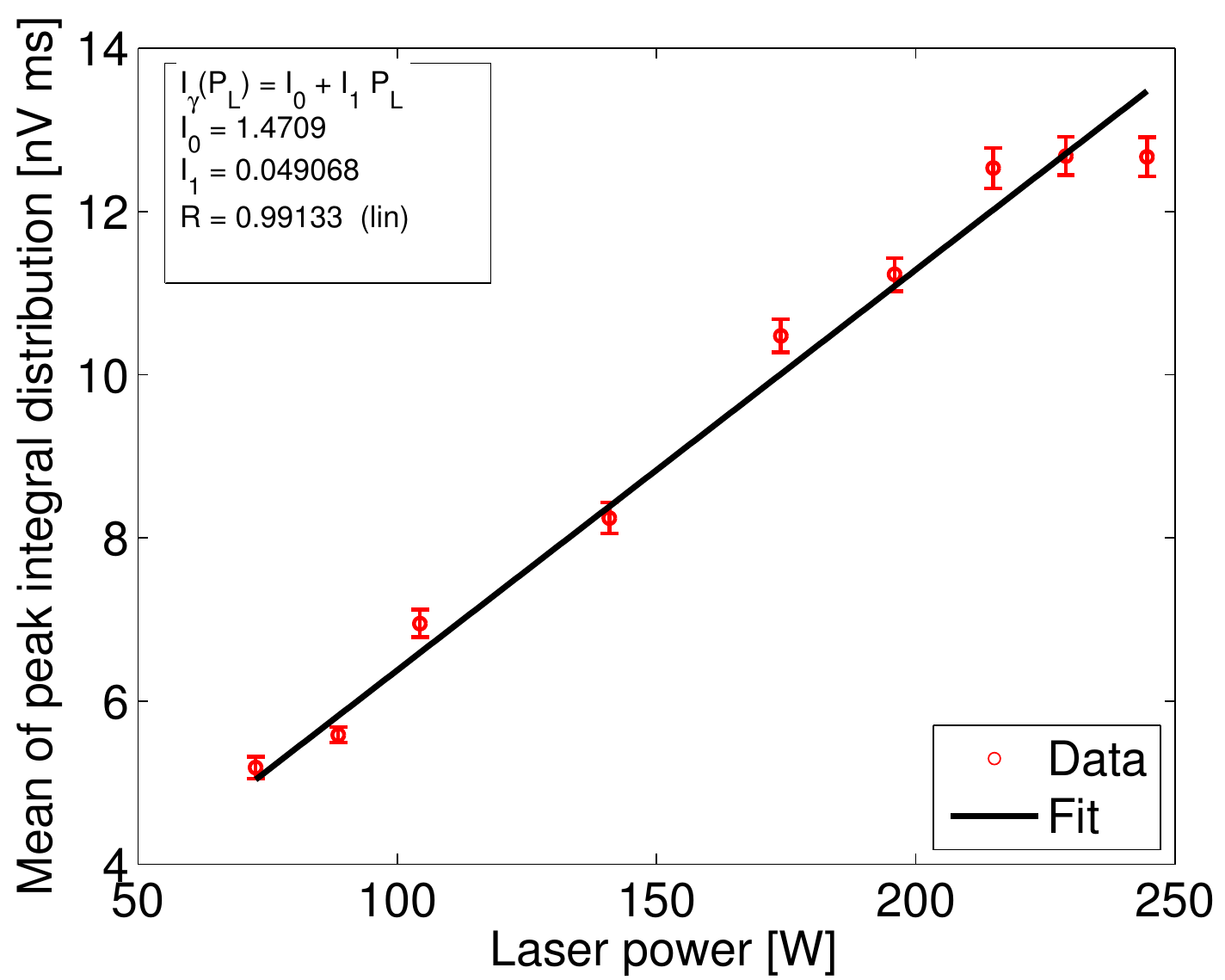}}
\caption{ Figure~\protect\ref{fig:mean_spectrum_PS} presents gamma spectrum for different FPC stored laser power. Different colors correspond to the spectra taken at different FPC powers. In the legend, the fraction of the maximum FPC stored power and the mean of each of these spectra are shown. Figure~\protect\ref{fig:spectrum_PS} shows the linear relation between the mean of gamma spectra for each laser power bin and the correspondent laser power which is defined as a mean of the laser power bins shown on figure~\protect\ref{fig:LP_histo}. The errors bars indicate the statistical errors.}
\label{fig:spectra_bins}
\end{figure}

To verify the linearity of the gamma ray flux as a function of the laser power stored in the FPC, we took data over a wide range of laser power (see subsection~\ref{par:LP}). The laser power distribution has been split into the nine bins of approximately 1300 events each as it is shown on figure~\ref{fig:LP_histo}.  Then, the gamma production has been studied within each laser power bin. 
The spectra of scattered gamma rays which correspond to the different values of  laser power stored in the FPC are presented on figure~\ref{fig:mean_spectrum_PS}. As expected, the average of the peak integral distribution (energy deposited in the calorimeter by the gamma rays) scales linearly with the laser power stored in the FPC (see figure~\ref{fig:spectrum_PS}).

\paragraph{Data - simulations comparison} 
Simulations were carried out to have a comparison with the gamma ray spectrum obtained experimentally .
Noise was introduced in the simulations and  quality cuts were applied in the same manner for  both data and  simulations.

\begin{figure}[htbp]
\begin{center}
\includegraphics[width=0.5\textwidth]{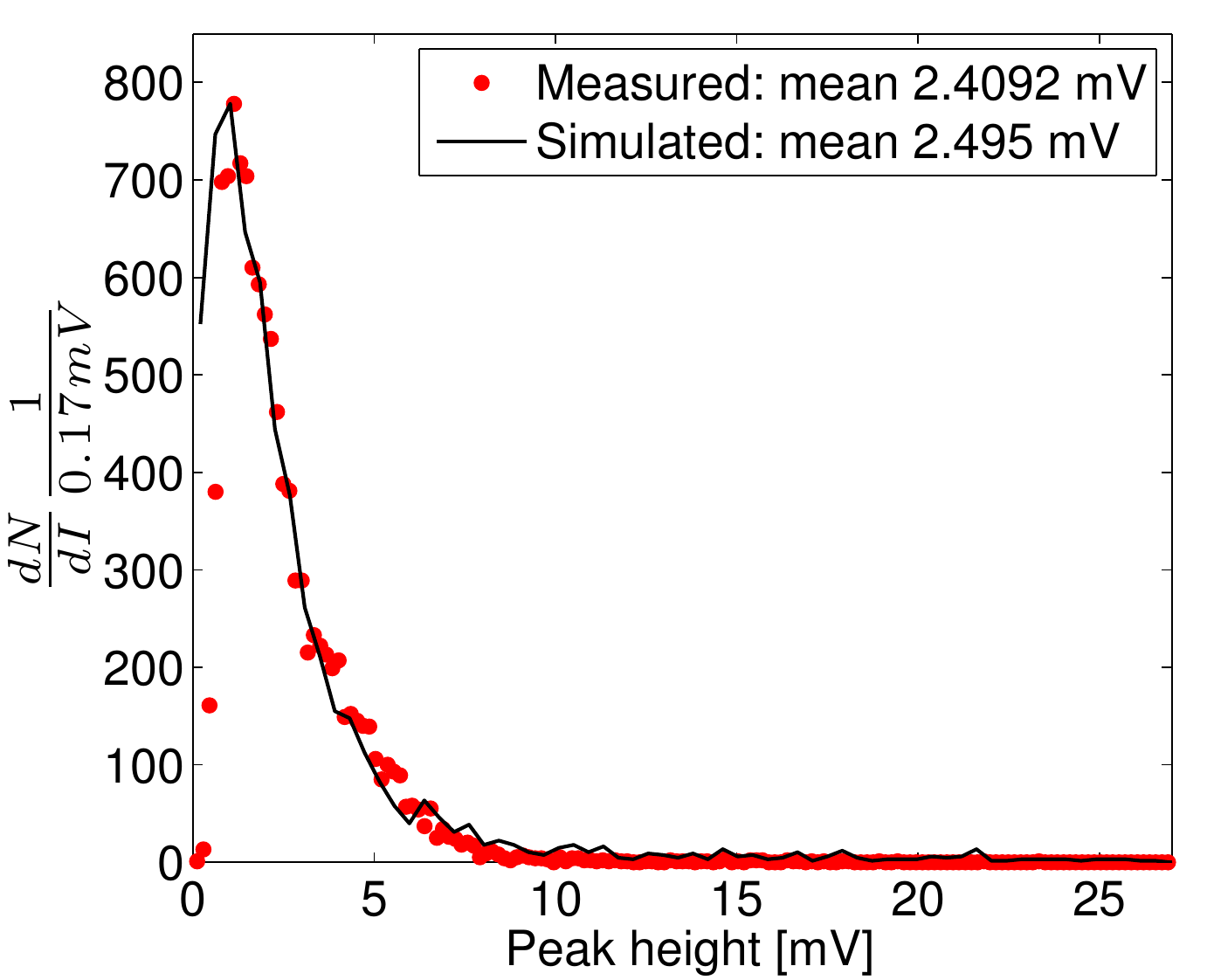}
\caption{Comparison between the measured and simulated energy spectra of the gamma rays resulting from Compton scattering. The black solid line represents the simulated spectrum and the circles dots represent the measured gamma ray energy distribution expressed in the peak heights. For the simulations the events occur according to a Poisson process with a rate of 2.7 gammas per bunch crossing. The data presented are the same than those of figure~\protect\ref{fig:HQ_PI}.}
\label{fig:data_simu}
\end{center}
\end{figure}

The spectrum of gamma rays measured is shown on figure~\ref{fig:data_simu} together with the spectrum simulated for a MPV of 2.7 gammas. The spectrum corresponds to the distribution of the energy deposited in the calorimeter expressed by the peak heights. 

\begin{figure}[!ht]
\begin{center}
\includegraphics[width=0.5\textwidth]{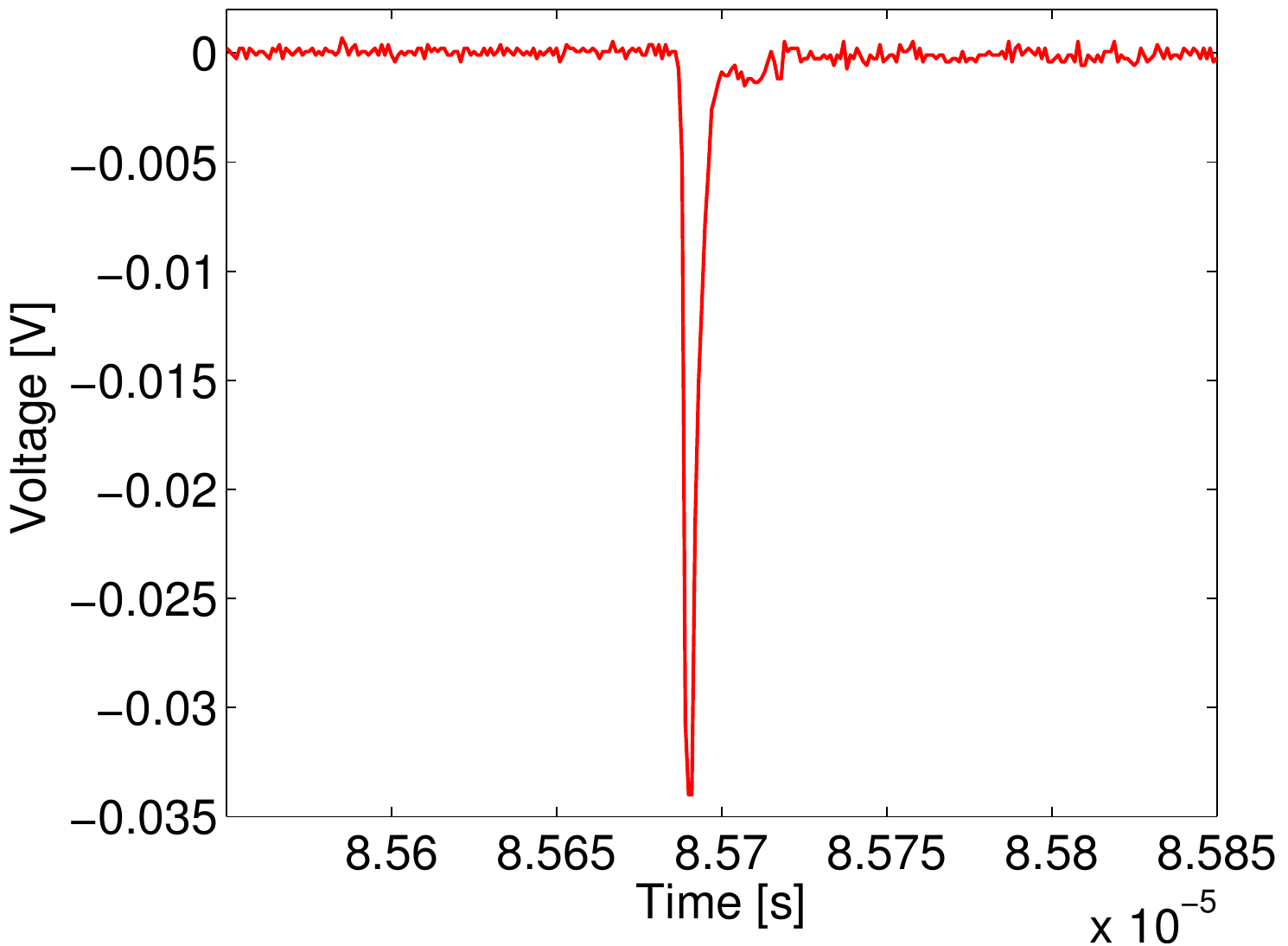}
\caption{Part of the data file containing the highest energy deposited in the calorimeter per one shot which is about 904~MeV. Waveform shows this event which stands for the highest instantaneous gamma flux we measured. It is equivalent to 38~gammas produced in one bunch crossing.}
\label{fig:BestInstFlux}
\end{center}
\end{figure}

Figure~\ref{fig:data_simu} shows a very good agreement between the experimental data and simulations. 
\paragraph{Highest integrated/instantaneous flux} 
We scanned our data for the best integrated and instantaneous gamma flux which was produced. The best instantaneous flux we measured (the highest energy deposition)  was~34~mV which according to the calibration corresponds to 904~$\pm$~65~MeV deposited in the calorimeter. Assuming~24~MeV per gamma, this gives about 38~$\pm$~3~gammas produced per bunch crossing. The corresponding Compton peak can be seen on figure~\ref{fig:BestInstFlux}.

\begin{figure}[H]
\centering
\subfloat[]{\label{fig:HIF1}\includegraphics[width=0.47\textwidth]{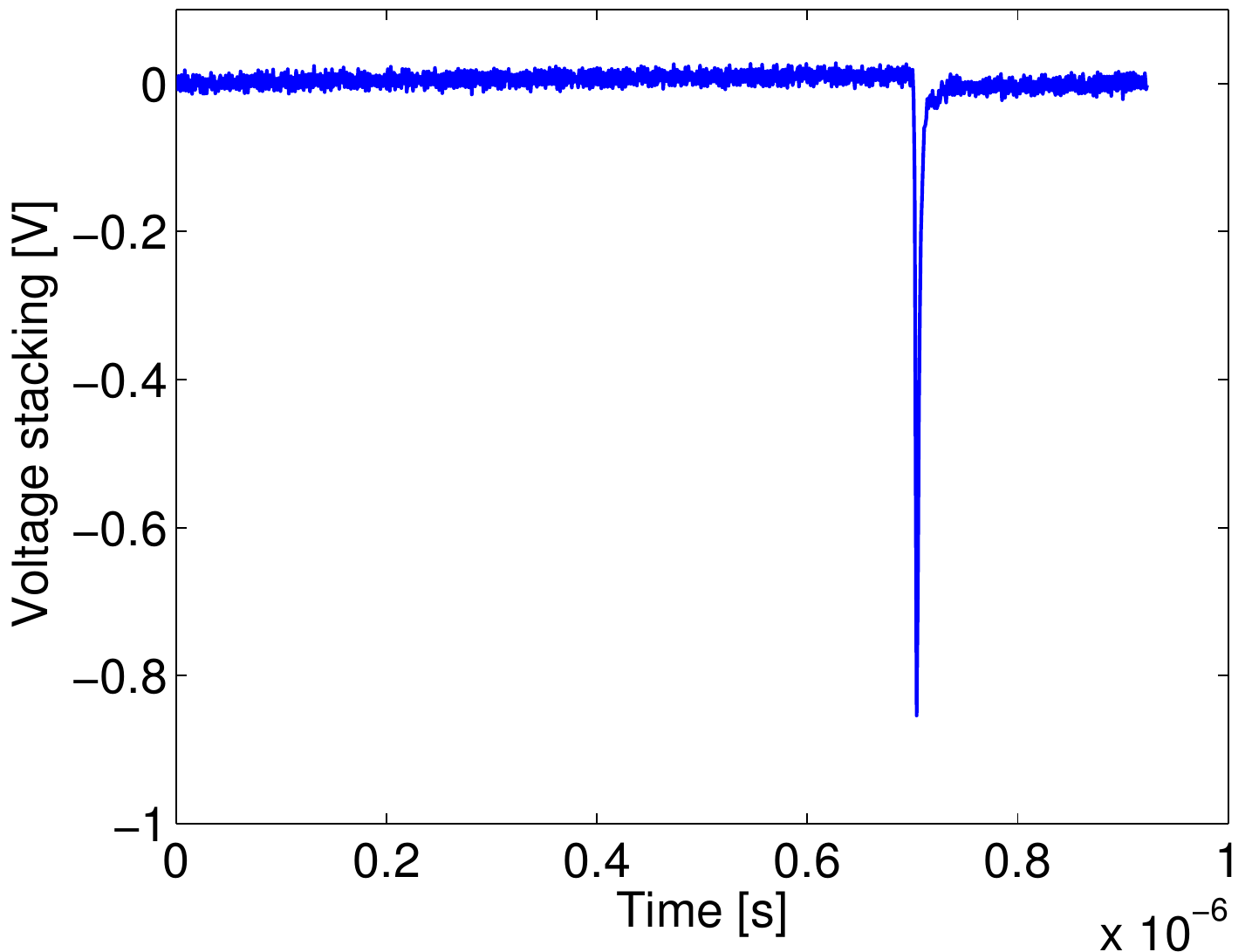}}   
\hspace{5mm}  
\subfloat[]{\label{fig:HIF2}\includegraphics[width=0.47\textwidth]{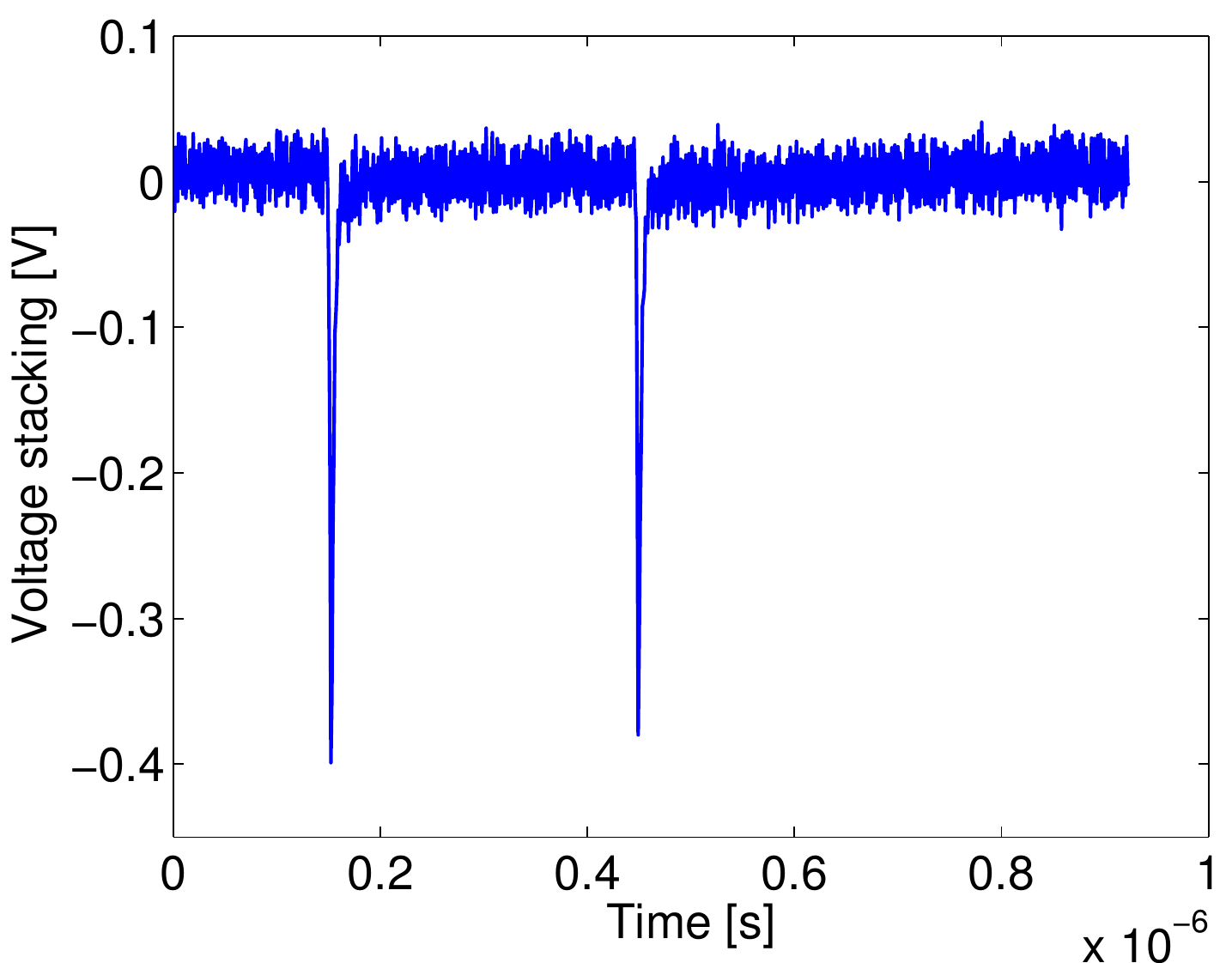}}
\hspace{5mm} 
\subfloat[]{\label{fig:HIF3}\includegraphics[width=0.47\textwidth]{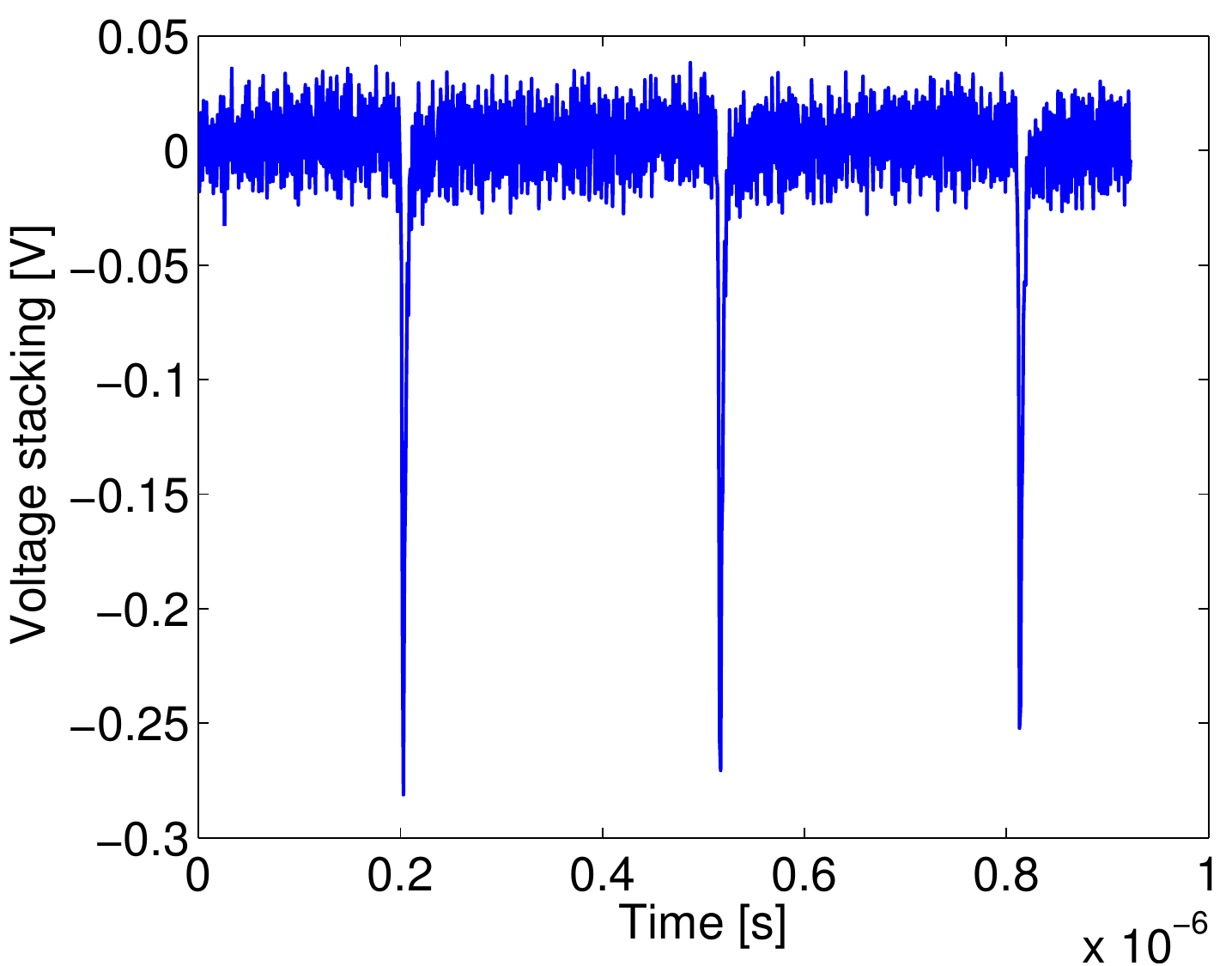}}
\hspace{5mm} 
\caption{Illustration of three data files containing the highest integrated flux over 0.2~ms for different ATF DR filling modes: one bunch stored in the DR (figure~\protect\ref{fig:HIF1}),  two trains of one bunch each stored in the DR (figure~\protect\ref{fig:HIF2}) and  three trains of one bunch each stored in the DR (figure~\protect\ref{fig:HIF3}). The figures show all the periods of a given file stacked together.  }
\label{fig:HIF}
\end{figure}

For the many applications of accelerator driven Compton sources an important characteristic is the total gamma flux. 
Table~\ref{table:bestFlux} summarizes the results concerning the highest  gamma ray flux integrated over 0.2~ms (duration of one data file) for different ATF DR filling modes.  As one can see, the total flux does not scale linearly with the number of electron bunches stored in the DR contrary to what was expected. Additional studies should be conducted to investigate this issue.
Figure~\ref{fig:HIF} shows the three data files having the best integrated gamma ray flux. 

\begin{table}[H]
\centering
\caption{Highest integrated gamma ray flux achieved.}
\begin{tabular}{|c|c|c|c|c|c|}
\hline
\small Electron pulse structure&\small  \begin{tabular}[x]{@{}c@{}}Total intensity\\over 0.2~ms\end{tabular} & \small  \begin{tabular}[x]{@{}c@{}}Energy deposited\\ over 0.2~ms\end{tabular}& \small  \begin{tabular}[x]{@{}c@{}} Integrated flux\\ over 0.2~ms\end{tabular}&  \small  \begin{tabular}[x]{@{}c@{}}  Integrated flux\\ over 1~s\end{tabular} & \begin{tabular}[x]{@{}c@{}}  Systematic \\ error \end{tabular} \\\hline \hline
\small 1 train &\small 893~mV& \small 23750 MeV & \small 990~ $\gamma$ & \small $\sim4.9\times10^6~ \gamma$ & 7\% \\ \hline
\small 2 trains &\small 910~mV & \small 24210 MeV& \small 1010~$\gamma$ & \small $\sim5.0\times10^6~ \gamma$& 7\%  \\ \hline
\small 3 trains &\small 1010~mV & \small 26800 MeV & \small 1120~ $\gamma$ & \small $\sim5.6\times10^6~ \gamma$& 7\%  \\ \hline
\end{tabular}
\label{table:bestFlux}
\end{table}

\newpage

\section{Conclusion and outlook}
\label{sec:concl}
The MightyLaser project has demonstrated the production of gamma rays using a non-planar~four-mirror  FPC. In average,  2.7~$\pm$~0.2~gammas are produced per bunch crossing. This result is in agreement with the CAIN simulations for the FPC laser power achieved in the experiment. A good agreement is also observed between the measured and simulated energy spectra of the gamma rays using Geant4 toolkit. The highest integrated flux of 1120~$\pm$~80~~gammas over 0.2~ms~(~$\sim~5.6~\times~10^6$~ gammas per second) in three trains mode and an instantaneous flux of 38~$\pm$~3~gammas per bunch crossing were measured.
These values are limited by the FPC laser power obtained so far.

Work to improve the laser stability and increase the FPC finesse has been interrupted by the earthquake which hit Japan in March 2011. Operation of the ATF at KEK is to be resumed soon. Our objective for the next campaign is to increase production of high energy gamma rays by several orders of magnitude by increasing both the FPC finesse and the input laser power.

\section*{Acknowledgements}

This work has been funded by the French ANR (Agence Nationale de la Recherche) contract number BLAN08-1-382932 and P2I (Physique des 2 Infinis).  The authors would like to thank the staff from KEK for their hospitality.

\bibliographystyle{unsrt}
\bibliography{biblio}

\begin{thebibliography}{10}

\bibitem{intrumentationPaper}
J.~Bonis et~al.
\newblock Non-planar four-mirror optical cavity for high intensity gamma ray
  flux production by pulsed laser beam compton scattering off
  {G}e{V}-electrons.
\newblock {\em To be submitted to JINST}.

\bibitem{araki2005design}
S.~Araki et~al.
\newblock Design of a polarised positron source based on laser compton
  scattering.
\newblock {\em Arxiv preprint physics/0509016}, 2005.

\bibitem{ILCpositrons}
M.~Kuriki et~al.
\newblock {ILC positron source based on laser Compton}.
\newblock {\em AIP Conf.Proc.}, 980:92--101, 2008.

\bibitem{CLICpositrons}
O.~Dadoun et~al.
\newblock {The Baseline Positron Production and Capture Scheme for CLIC}.
\newblock 1st International Particle Accelerator Conference: IPAC'10, 23-28 May
  2010, Kyoto, Japan.

\bibitem{louvre}
P.~Walter et~al.
\newblock A new high quality x-ray source for cultural heritage.
\newblock {\em Comptes Rendus Physique}, 10(7):676 -- 690, 2009.

\bibitem{ThomX}
ThomX collaboration.
\newblock Thomx {CDR}.
\newblock {\em IN2P3}, in2p3-00448278, 2010.

\bibitem{ThomX2}
A.~Loulergue et~al.
\newblock {A Compact Ring for the ThomX-ray Source}.
\newblock 1st International Particle Accelerator Conference: IPAC'10, 23-28 May
  2010, Kyoto, Japan.

\bibitem{quantumbeam}
J.~Urakawa.
\newblock Development of a compact x-ray source based on compton scattering
  using a 1.3 {GH}z superconducting {RF} accelerating linac and a new laser
  storage cavity.
\newblock {\em Nuclear Instruments and Methods in Physics Research Section A:
  Accelerators, Spectrometers, Detectors and Associated Equipment}, 2010.

\bibitem{compactref}
P.~Sprangle et~al.
\newblock Tunable, short pulse hard x-rays from compact laser synchrotron
  source.
\newblock {\em Appl. Phys.}, 72:5032--5034, 1992.

\bibitem{kogel}
H.~Kogelnick and T.~Li.
\newblock Laser beams and resonators.
\newblock {\em Appl. Opt.}, 5:1550--1567, 1966.

\bibitem{ruth}
Z.~Huang and R.~D. Ruth.
\newblock Laser-electron storage ring.
\newblock {\em Phys. Rev. Lett.}, 80(5):976--979, Feb 1998.

\bibitem{ATF}
F.~Hinode et~al.
\newblock {ATF} accelerator test facility design and study report {N}o. 4.
  {KEK}, {T}sukuba, {J}apan, 1995.
\newblock \url{http://lcdev.kek.jp/ATF/Pub/KEK-I-95-4.pdf}.

\bibitem{shimizu2009photon}
H.~Shimizu et~al.
\newblock Photon generation by laser-compton scattering using an optical
  resonant cavity at the {KEK-ATF} electron ring.
\newblock {\em Journal of the Physical Society of Japan}, 78(7):4501, 2009.

\bibitem{miyoshi2010photon}
S.~Miyoshi et~al.
\newblock Photon generation by laser-compton scattering at the {KEK-ATF}.
\newblock {\em Nuclear Instruments and Methods in Physics Research Section A:
  Accelerators, Spectrometers, Detectors and Associated Equipment}, 2010.

\bibitem{ATF-2000}
H.Hayano et~al.
\newblock Accelerator test facility study report {JFY} 19996-1999.
\newblock {\em KEK, Tsukuba, Japan}, KEK Internal report 2000-6, 2000.

\bibitem{ATF2}
P.~Bambade et~al.
\newblock Present status and first results of the final focus beam line at the
  {KEK} {A}ccelerator {T}est {F}acility.
\newblock {\em Phys. Rev. ST Accel. Beams}, 13(4):042801, Apr 2010.

\bibitem{ATF-Timing}
T.Naito et~al.
\newblock Timing system for multi-bunch/multi-train operation at {ATF-DR}.
\newblock {\em KEK, Tsukuba, Japan}, KEK Preprint 99-144, 1999.

\bibitem{Miyoshithesis}
S.~Miyoshi.
\newblock {\em Development of polarized positron source by laser {C}ompton
  scattering with optical resonant cavity}.
\newblock PhD thesis, Graduate School of Advanced Sciences of Matter, Hiroshima
  University, 2011.

\bibitem{GEANT4}
S.~Agostinelli et~al.
\newblock Geant4: A simulation toolkit.
\newblock {\em Nucl. Instrum. Meth.}, A506:250--303, 2003.

\bibitem{cain}
K.~Yokoya.
\newblock User manual of {CAIN}, version 2.40, 2009.

\bibitem{suzukigeneral}
T.~Suzuki.
\newblock General formulae of luminosity for various types of colliding beam
  machines.
\newblock {\em KEK note}, pages 76--3.

\bibitem{landau}
V.~B. Berestetskii, L.~D. Landau, E.~M. Lifshitz, L.~P. Pitaevskii, J.~B.
  Sykes, and J.~S. Bell.
\newblock {\em {Quantum electrodynamics}}, volume~4.
\newblock Butterworth-Heinemann, 1982.

\bibitem{landau1944energy}
L.~Landau.
\newblock On the energy loss of fast particles by ionisation.
\newblock {\em J. Phys. {USSR}}, 8(4):201, 1944.

\bibitem{groom2011atomic}
D.~Groom.
\newblock Atomic and nuclear properties of materials. {P}article {D}ata {B}ook.
\newblock \url{http://pdg.lbl.gov/2011/AtomicNuclearProperties/}.

\end{thebibliography}

\end{document}